\documentclass[a4paper, 10pt]{article}

\usepackage{draft}
\usepackage{mathbbol}
\usepackage[mathscr]{eucal}

\newcommand{\InEq}[1]{\qquad\text{#1}\qquad}
\newcommand{\Iff}{\qquad\Longleftrightarrow\qquad}
\newcommand{\Implies}{\qquad\Longrightarrow\qquad}

\newcommand{\smf}{\sf\scriptscriptstyle}
\newcommand{\smbr}[1]{\scriptscriptstyle[#1]}
\newcommand{\AmbInd}[1]{\mathsf{#1}}
\newcommand{\Base}{\mathscr{X}}
\newcommand{\dR}{\mathrm{d}}
\newcommand{\End}{\mathrm{End}}
\newcommand{\Forms}{{\bigwedge}}
\newcommand{\Functions}{\mathscr{C}^\infty}
\newcommand{\Ker}{\mathrm{Ker}}
\newcommand{\tdeg}{\mathrm{tdeg}}

\newcommand{\cE}{\mathcal{E}}
\newcommand{\cV}{\mathcal{V}}
\newcommand{\1}{\mathbf{1}}
\newcommand{\gMfd}[1]{\mathcal{#1}}
\newcommand{\Q}{\mathcal{Q}}
\newcommand{\C}{\mathbb{C}}
\newcommand{\N}{\mathbb{N}}
\newcommand{\R}{\mathbb{R}}
\newcommand{\Z}{\mathbb{Z}}
\newcommand{\g}{\mathfrak{g}}
\newcommand{\h}{\mathfrak{h}}

\newcommand{\hs}{\mathfrak{hs}}
\newcommand{\iso}{\mathfrak{iso}}
\newcommand{\so}{\mathfrak{so}}
\renewcommand{\sl}{\mathfrak{sl}}
\newcommand{\su}{\mathfrak{su}}
\newcommand{\cC}{\mathcal{C}}
\newcommand{\cI}{\mathcal{I}}

\newcommand{\gh}[1]{\mathrm{gh}(#1)}

\newcommand{\dv}{\mathrm{d_v}}
\newcommand{\dx}{d_\Base}

\renewcommand{\dh}{\mathrm{d_h}}

\newcommand\blfootnote[1]{%
  \begingroup
  \renewcommand\thefootnote{}\footnote{#1}%
  \addtocounter{footnote}{-1}%
  \endgroup
}
\newcommand{\fud}[2]{{}^{#1}{}_{#2}\,}

\begin{document}

\title{%
    Multisymplectic \\
    {\color{MidnightBlue}\adfopenflourishleft}\,
    \raisebox{1.5pt}{%
        \color{PineGreen}$\scriptscriptstyle\blacklozenge$%
    }\,
    {\color{MidnightBlue}\adfopenflourishright} \\
    AKSZ sigma models
}
\author{Thomas {\sc Basile},}
\author{Maxim {\sc Grigoriev$^{\dagger,\ddagger}$}\quad\&}
\author{Evgeny {\sc Skvortsov$^{\dagger}$}}
\affiliation{%
    Service de Physique de l’Univers,
    Champs et Gravitation,\\
    Universit\'e de Mons, 20 place du Parc,
    7000 Mons, Belgium%
}
\emailAdd{thomas.basile@umons.ac.be}
\emailAdd{maxim.grigoriev@umons.ac.be}
\emailAdd{evgeny.skvortsov@umons.ac.be}
\abstract{%
    The Alexandrov--Kontsevich--Schwarz--Zaboronsky 
    (AKSZ) construction encodes all the data of a topological
    sigma-model in the finite-dimensional symplectic $Q$-manifold. 
    Relaxing the nondegeneracy condition i.e. considering
    a presymplectic form instead, extends the construction
    to non-topological models. The gauge-invariant action functional 
    of (presymplectic) AKSZ sigma model is written in terms
    of space-time differential forms and can be seen as a covariant 
    multidimensional analogue of the usual 1st order
    Hamiltonian action. In this work, we show that the AKSZ 
    construction has a natural generalisation where the target space 
    $Q$-manifold is equipped with a form of arbitrary degree
    $\Omega$ (possibly inhomogeneous) which is $(\mathrm{d}+L_Q)$-closed. This data defines
    a higher-derivative generalisation of the AKSZ action
    which is still invariant under the natural gauge transformations 
    determined by $Q$ and which is efficiently formulated in terms
    of a version of Chern--Weil map introduced by Kotov and Strobl. It turns out that a variety of interesting
    gauge theories, including higher-dimensional Chern--Simons 
    theory, MacDowell--Mansouri--Stelle--West action and self-dual gravity as well as its higher spin extension,
    can be concisely reformulated as such multisymplectic
    AKSZ models. We also present a version of the construction
    in the setup of PDE geometry and demonstrate that the counterpart 
    of the multisymplectic AKSZ action is precisely 
    the standard multisymplectic formulation, 
    where the Chern--Weil map corresponds to the usual pullback map. 
}

\maketitle

\blfootnote{${}^{\dagger}$ Also at Lebedev Physical Institute, Moscow, Russia.}

\blfootnote{${}^{\ddagger}$ Also at Institute for Theoretical and Mathematical Physics, Lomonosov MSU, Moscow, Russia. }

\newpage
\section{Introduction}
AKSZ sigma models, dubbed after Alexandrov, Kontsevich,
Schwarz and Zaboronsky \cite{Alexandrov:1995kv} are \emph{topological}  sigma models which can be defined
out of relatively simple data, namely that of a \emph{symplectic} 
$Q$-manifold.\footnote{The literature on AKSZ sigma models is very extensive so we mention only the very early relevant contributions~\cite{Cattaneo:1999fm,Grigoriev:1999qz,Cattaneo:2001ys,Batalin:2001fc,Roytenberg:2002nu,Hofman:2002rv}. For a review and further references see e.g.~\cite{Ikeda:2012pv}.
} 
More precisely, one asks that this manifold $\gMfd{M}$
be equipped with a homological vector field 
i.e. a self-commuting vector field of degree $1$,
\begin{equation}
    Q \in \Gamma(T\gMfd{M})\,,
    \qquad 
    \deg(Q) = 1\,,
    \qquad 
    Q^2 \equiv \tfrac12[Q,Q]=0\,,
\end{equation}
as well as a \emph{closed} and \emph{non-degenerate} $2$-form
$\omega \in \bigwedge^2 \gMfd{M}$, $\deg(\omega)=n-1$ preserved by
the homological vector field,
\begin{equation}
    \dR\omega = 0 = L_Q\omega\,.
\end{equation}
For $n \neq 1$, the closed $2$-form $\omega$
is necessarily exact, 
\begin{equation}
    \omega = \dR\chi\,,
    \InEq{for} \Theta \in \Forms^1 \gMfd{M}\,,
    \qquad \deg(\Theta) = n-1\,,
\end{equation}
and if $n\neq 0$, the homological vector field $Q$
admits a global Hamiltonian, meaning
\begin{equation}
    \imath_Q\omega + \dR\mathcal{L} = 0\,,
    \InEq{for} \mathcal{L} \in \Functions(\gMfd{M})\,,
    \quad \deg(\mathcal{L}) = n\,,
\end{equation}
as a consequence of the compatibility between $\omega$ and $Q$
(see e.g. \cite{Roytenberg:2002nu}). Given this data,
one defines a topological sigma model on an $n$-dimensional
smooth manifold $\Base$ via
\begin{equation}\label{eq:AKSZ}
    S[\varphi] = \int_{T[1]\Base} 
    \Big(\imath_{d_\Base}\varphi^*\chi
        + \varphi^*\mathcal{L}\Big)\,,
    \qquad
    T[1]\Base \overset{\varphi}{\longrightarrow} \gMfd{M}\,,
\end{equation}
where $d_\Base$ denotes the homological vector field on $T[1]\Base$
corresponding to the de Rham differential on $\Base$.
As suggested by the notation above, the fields of this sigma model 
are maps between $T[1]\Base$ and $\gMfd{M}$, with the following
added bonus:  while the space of degree-preserving maps,
which is the usual  notion of morphisms
between ($\Z$-)graded manifolds, 
accounts for the \emph{classical fields} of the theory
(of ghost number $0$), the space of all maps---degree-preserving
or not---corresponds to the space of classical fields,
extended with ghosts, and their antifields. In other words,
relaxing the condition that the maps between the source
and target graded manifolds be degree-preserving%
\footnote{Such, non-degree-preserving, maps
are often called `super-maps'.} yields the field content
necessary for the Batalin--Vilkovisky (BV) formulation
of the sigma model.

One of the simplest example of AKSZ models is Chern--Simons theory,
for which the relevant target space is the degree $2$ symplectic
$Q$-manifold given by the suspension $\g[1]$
of a \emph{quadratic} Lie algebra, that is a Lie algebra $\g$
equipped with a symmetric, non-degenerate and ad-invariant
bilinear form.\footnote{One should in fact refine slightly
this statement: the target manifold $\g[1]$ only capture
the case of trivial $G$-bundle.}
More concretely, denoting the degree $1$ coordinates on $\g[1]$
by $c^a$ with $a=1,\dots,\dim\g$, the symplectic form
$\omega$, and homological vector field $Q$, is given by
\begin{equation}\label{eq:g[1]}
    \omega = \tfrac12 \omega_{ab}\,\dR c^a \dR c^b\,,
    \qquad 
    Q = -\tfrac12\,f_{ab}{}^c\, c^a c^b\,
    \tfrac{\partial}{\partial c^c}\,,
\end{equation}
where $\omega_{ab}$ and $f_{ab}{}^c$ are the components
of the invariant bilinear form and the structure constants,
respectively, of $\g$ expressed in the basis dual
to the chosen coordinate system on $\g[1]$.
That the vector field  $Q$ squares to zero follows 
straightforwardly from the Jacobi identity for the Lie bracket 
of $\g$ and the symplectic form $\omega$ is obviously closed. 
The compatibility between $Q$ and $\omega$ reads
\begin{equation}
    L_Q\omega = 0
    \Iff
    \omega_{d(a}f_{b)c}{}^d = 0
\end{equation}
which is the ad-invariance condition for $\omega$, i.e.
\begin{equation}
    \omega\big([x,y]_\g,z\big) = \omega\big(x,[y,z]_\g\big)\,,
    \qquad \forall x,y,z \in \g\,.
\end{equation}
The $1$-form potential for $\omega$ and Hamiltonian function
for $Q$ are easily found to be
\begin{equation}
    \chi = \tfrac12 \omega_{ab}\, c^a \dR c^b\,,
    \qquad 
    \mathcal{L}
        = \tfrac16\,\omega_{ad} f_{bc}{}^d\, c^a c^b c^c\,,
\end{equation}
so that, put together one ends up recovering the well-known
Chern--Simons action functional,
\begin{equation}
    S_{\smf CS}[A]
    = \tfrac12\,\int_\Base \omega\big(A, \dR A
    + \tfrac13[A,A]_\g\big)\,.
\end{equation}
As it turns out, the above Chern--Simons functional
is only the first of a family of topological theory
that can be defined in \emph{odd dimensions}, as the integral
of \emph{higher Chern--Simons terms}. Given a symmetric,
non-degenerate and ad-invariant multilinear form of arity $p$
on the Lie algebra $\g$, say $\Omega \in (S^p\g^*)^\g$,
the latter are defined as the $(2p-1)$-form
$\mathsf{CS}(A) \in \Forms^{2p-1} \Base$,
made out of the $\g$-valued gauge field $A$, via
\begin{equation}\label{eq:CS_form}
    \Omega(\underbrace{F,\dots,F}_{p\ \text{times}})
    = \dR \mathsf{CS}(A)\,,
    \InEq{with}
    F[A] = \dR A + \tfrac12 [A,A]_\g\,.
\end{equation}
From the point of view of the graded manifold $\g[1]$,
the invariant polynomial is a \emph{closed} $p$-form
preserved by $Q$, 
\begin{equation}
    \Forms^p \g[1]\ \ni\ \Omega
    = \tfrac1{p!}\Omega_{a_1 \dots a_p}
        \dR c^{a_1} \dots \dR c^{a_p}\,\,,
    \qquad 
    \dR\Omega = 0 = L_Q\Omega\,.
\end{equation}
If this closed $p$-form also verifies the non-degeneracy
condition
\begin{equation}
    \imath_X\Omega = 0\,,
    \qquad X \in \Gamma(T\g[1]) \Implies X = 0\,,
\end{equation}
then it defines a $Q$-invariant \emph{multisymplectic} form
on $\g[1]$. This begs the question: can we think of Chern--Simons
theory in higher dimensions as resulting from
a \emph{multisymplectic} version of the AKSZ construction?
If so, what are the details and possible applications
of such a construction?

The paper is organized as follows: in Section \ref{sec:multi_AKSZ} we recall the notion of Chern--Weil map for $Q$-bundles, initially put forward in~\cite{Kotov:2007nr} and reformulate the AKSZ construction and its variations in such terms. Then we define multisymplectic AKSZ sigma-models and give a description of gauge symmetries. Section \ref{sec:examples} is devoted to the examples which include higher-dimensional Chern--Simons theory, MacDowell--Mansouri--Stelle--West gravity and self-dual gravity as well as its higher spin extension. Finally, in Section \ref{sec:PDE} we explicitly relate the multisymplectic AKSZ to conventional multisymplectic formalism focusing on the case where underlying bundle is a partial differential equation (PDE).

\section{Multisymplectic AKSZ sigma models}
\label{sec:multi_AKSZ}

\paragraph{Chern--Weil morphism for $Q$-manifolds.}
Let us consider a pair of $Q$-manifolds $(\gMfd{X}, d_\gMfd{X})$
and $(\gMfd{M},Q)$, and a morphism of graded manifolds
$\sigma:\gMfd{X} \longrightarrow \gMfd{M}$ between them. As shown in \cite{Kotov:2007nr}, the pullback of such a morphism can be lifted to a morphism of differential graded
commutative algebras,
\begin{equation}
    \sigma^*_W: \big(\Forms^\bullet\gMfd{M}, \dR+L_Q\big)
    \longrightarrow \big(\Functions(\gMfd{X}), d_\gMfd{X}\big)\,.
\end{equation}
It is uniquely determined by its action on functions and exact 1-forms via:
\begin{equation}
    \sigma^*_W(f) = \sigma^*(f)\,,
    \qquad 
    \sigma^*_W(\dR f)
        = d_\gMfd{X}\sigma^*(f) - \sigma^*\big(Qf\big)\,.
\end{equation}
and is extended to generic forms as an algebra morphism. In practice, we define it for $\gMfd{M}$ coordinates $z^I$ and the corresponding
$1$-form basis $\dR z^I$:
\begin{equation}
    \sigma^*_W(z^I) = \sigma^*(z^I)\,,
    \qquad 
    \sigma^*_W(\dR z^I)
        = d_\gMfd{X}\sigma^*(z^I) - \sigma^*\big(Q^I(z)\big)\,.
\end{equation}
The relation between AKSZ sigma models and structures known in the context of characteristic classes and equivariant cohomology was discussed in e.g.~\cite{Fiorenza:2011jr}. Let us also mention that characteristic classes of $Q$-manifolds have been studied in~\cite{Lyakhovich:2004kr,Lyakhovich:2009qq}.

It is useful to explicitly express the morphism $\sigma^*_W$ in terms of the underlying morphism $\sigma^*$:
\begin{equation}
    \sigma^*_W = p \circ e^{\imath_{d_\gMfd{X}}} \circ
        \sigma^* \circ e^{-\imath_Q}\,,
    \InEq{where}
    p: \Forms^{\bullet}\gMfd{X} \twoheadrightarrow \Functions(\gMfd{X})\,,
\end{equation}
i.e. map $p$ is nothing but a projection of a form
on $\gMfd{X}$ onto its $0$-form component. Indeed, one can easily check that the above expression reproduces \eqref{eq:CW} when evaluated on $0$-forms and exact $1$-forms on $\gMfd{M}$, and is a morphism of algebra. The latter property follows from the fact that $\imath_Q$ and $\imath_{d_\gMfd{X}}$ are (degree $0$) derivations of $\Forms^{\bullet}\gMfd{E}$ and $\Forms^{\bullet}\gMfd{X}$ respectively, and that the exponential of a derivation of a commutative algebra is an algebra morphism.

In order to check that the previous formula defines
a chain map, we should first note that the `twist'
of the de Rham differential on $\gMfd{M}$ by the Lie derivative along $Q$ can be understood as resulting
from the transfer of the former under the isomorphism
$e^{-\imath_Q}$, that is%
\begin{equation}\label{eq:exp}
    e^{-\imath_Q}
        \circ \big(\dR + L_Q\big)
        = \dR \circ e^{-\imath_Q}\,.
\end{equation}
Put differently, 
\begin{equation}\label{eq:iso_Weil}
    \begin{tikzcd}[column sep=large]
    \big(\Forms^\bullet\gMfd{M}, \dR + L_Q\big) 
    \quad \ar[r, "e^{-\imath_Q}"]
    & \quad \big(\Forms^\bullet\gMfd{M}, \dR\big)
    \end{tikzcd}
\end{equation}
is an isomorphism of differential graded commutative algebras,
where on both sides 
the grading is taken to be the \emph{total degree}
that we shall denote $\tdeg(-)$,
i.e. the sum of form degree and homological degree.
Using this simple identity, one finds
\begin{subequations}
\begin{align}
    e^{\imath_{d_\gMfd{X}}} \circ \sigma^* \circ e^{-\imath_Q}
    \circ \big(\dR + L_Q\big)
    & = e^{\imath_{d_\gMfd{X}}} \circ \sigma^* \circ 
    \dR \circ e^{-\imath_Q} \\ 
    & = e^{\imath_{d_\gMfd{X}}} \circ \dR \circ \sigma^*
    \circ e^{-\imath_Q} \\ 
    & = \big(\dR +L_{d_\gMfd{X}}\big) \circ e^{\imath_{d_\gMfd{X}}}
    \circ \sigma^* \circ e^{-\imath_Q}\,,
\end{align}
\end{subequations}
which, combined with 
\begin{equation}\label{eq:proj}
    p \circ \big(\dR + L_{d_\gMfd{X}}) = p \circ L_{d_\gMfd{X}}
    = d_\gMfd{X} \circ p\,,
\end{equation}
leads to
\begin{equation}
    \sigma_W^* \circ \big(\dR + L_Q\big)
    = d_\gMfd{X} \circ \sigma_W^*\,,
\end{equation}
as claimed initially. To summarise, the Chern--Weil morphism
is a composition of maps,
\begin{equation*}
    \begin{tikzcd}
        \big(\Forms^\bullet\gMfd{M}, \dR+L_Q\big)
        \ar[r, "e^{-\imath_Q}"]
        & \big(\Forms^\bullet\gMfd{M}, \dR\big)
        \ar[r, "\sigma^*"]
        & \big(\Forms^\bullet\gMfd{X}, \dR\big)
        \ar[r, "e^{\imath_{d_\gMfd{X}}}"]
        & \big(\Forms^\bullet\gMfd{X}, \dR+L_{d_\gMfd{X}}\big)
        \ar[r, "p"]
        & \big(\Functions(\gMfd{X}), d_\gMfd{X}\big)\,,
    \end{tikzcd}
\end{equation*}
all of which are morphisms of \emph{differential graded commutative
algebras}.

In this paper, we are solely concerned with the case $\gMfd{X}=T[1]\Base$ equipped with its canonical homological vector field $d_\gMfd{X}=d_\Base$
corresponding to the de Rham differential on $\Base$.\footnote{Note that the generalisation to the case where $\gMfd{X}$ is itself a $Q$-bundle over $T[1]\Base$ is straightforward and corresponds to the explicit incorporation of background fields, see~\cite{Dneprov:2025eoi} for the construction in the presymplectic.} Each target space coordinate $z^I$ gives rise to a space time (gauge) field $A^I \equiv \sigma^*(z^I)$ which is naturally interpreted as a differential form of degree $k=\deg{z^I}$ on $\Base$. Note that negative degree coordinates do not have their associated fields. At the same time, applying the Chern--Weil map to $\dR z^I$ gives:
\begin{equation}\label{eq:CW}
    \sigma_W^*(\dR z^I) = F^I(A)
        \equiv d_\gMfd{X} A^I - Q^I(A)\,.
\end{equation}
The functions $F^I$ are components of the curvature of $\sigma$. 
Their field-theoretical meaning is that these are the components
of the equations of motion of the underlying non-Lagrangian
AKSZ sigma model, i.e. the model whose fields are maps
$T[1]\Base \longrightarrow \gMfd{M}$ and solutions are those maps 
that preserve the $Q$-structures. Note that $F^I$ can be nonvanishing 
if the degree of $z^I$ is either $0$ or $1$. In the latter case
the term with $d_\gMfd{X} A^I$ vanishes while $Q^I(A)$
generally does not. In other words, degree $-1$ coordinates
encoded constraints (=algebraic equations) as was originally
observed in~\cite{Barnich:2006hbb}.

\paragraph{Multisymplectic AKSZ actions.}
Suppose that the $Q$-manifold $\big(\gMfd{X},d_\gMfd{X}\big)$
admits a $d_\gMfd{X}$-invariant measure of degree $-n$, that is
\begin{equation}
    \int_\gMfd{X}(-): \Functions(\gMfd{X}) \longrightarrow \R\,,
    \InEq{such that}
    \int_\gMfd{X} d_\gMfd{X} f = 0\,,
    \qquad \forall f \in \Functions(\gMfd{X})\,.
\end{equation}
For instance, if $\gMfd{X}=T[1]\Base$ with $\Base$ an ordinary smooth manifold of dimension $\dim\Base=n$ without boundary, the Berezin integral defines such an operation,
with the invariance property being nothing but
the incarnation of Stokes theorem in $\Base$. More generally, assuming that functions on $\gMfd{X}$ are sections of a vector bundle over
a real manifold $\Base$ and the product is linear over $\cC^\infty(\gMfd{X})$ one can require that for any $f \in \Functions(\gMfd{X})$:
\begin{equation}
 \int_\gMfd{X} d_\gMfd{X} f = \int_\Base  \alpha_f\,, 
    \qquad \alpha_f = d \beta\,.
\end{equation}
Here, $\alpha_f$ is a density on $\Base$ which is, roughly speaking, obtained by integrating over the fibre coordinates and then represented as a top-form on $\Base$.\footnote{Under these conditions one can even  replace the algebra of function on $T[1]\Base$ with a more general differential graded algebra, e.g. not freely-generated, see~\cite{Bonechi:2009kx,Costello:2011,Bonechi:2022aji,Grigoriev:2025vsl}.} In this way implements the usual setup for boundary terms.

Let $\Omega \in \Forms^\bullet\gMfd{M}$
be a form of total degree $n+1$, such that
\begin{equation}
    \big(\dR + L_Q\big)\Omega = 0\,.
\end{equation}
In light of the isomorphism \eqref{eq:iso_Weil},
the cohomology of $\dR+L_Q$ is isomorphic
to the de Rham cohomology of $\gMfd{M}$,
which is concentrated in homological degree $0$,
see e.g.~\cite{Roytenberg:2006qz} (as is the case for all $\Z$-graded manifold). In particular, this means that if $\Omega$ has non-zero homological degree, then there exists a form $\Theta_W \in \Forms\gMfd{M}$ of total degree $n$, such that
\begin{equation}\label{eq:exact_form}
    \Omega = \big(\dR + L_Q\big)\Theta_W
    \Iff 
    \dR\Theta = e^{-\imath_Q}\Omega\,,
    \InEq{with}
    \Theta = e^{-\imath_Q}\,\Theta_W\,.
\end{equation}
The closed form $\Omega_W=e^{-\imath_Q}\Omega$ may be degenerate and, indeed, it is the case in some of the examples we consider below. That being said, we will refer to such models as multisymplectic whereas a more precise term would be pre-multisymplectic. 
The inhomogeneous form $\Theta$ can then be used 
to define an action principle, via
\begin{equation}\label{eq:action_multi-AKSZ}
    S[\sigma] = \int_{\gMfd{X}} \sigma^*_W(\Theta_W)
    = \int_{\gMfd{X}} p\big(e^{\imath_{d_\gMfd{X}}}\,
    \sigma^*\Theta\big) \equiv \int_{\gMfd{X}} 
    p\bigg(\sum_{k=0}^{p-1} \tfrac{1}{k!}\,
    \underbrace{\imath_{d_\gMfd{X}} \dots \imath_{d_\gMfd{X}}}_{k}
    \sigma^* \Theta_{\smbr{k}}\bigg)\,,
\end{equation}
where $\sigma: \gMfd{X} \longrightarrow \gMfd{M}$
is a (degree-preserving) map of graded manifolds,
$\Theta_{\smbr{k}} \in \Forms^k\gMfd{E}$ denotes
the $k$-form component of the inhomogeneous form $\Theta$,
and $p$ is the highest form degree of the components of $\Omega$.
In plain word, the action is nothing but the integral
over $\gMfd{X}$ of the pullback by the Chern--Weil morphism
of the inhomogeneous form $\Theta_W$ 
such that $(\dR+L_Q)\Theta_W=\Omega$.

\paragraph{Reformulation in terms of $Q$-bundles.}
Even if we limit ourselves to the usual sigma-model setup
where the space of fields is the space of maps from the source space 
to the target space it is very convenient to represent fields
as sections of the trivial bundle whose fibre is the target space. 
This setup also admits a natural generalisation where the bundle
is not necessarily trivial and the structures therein
are not factorised. In the context of $Q$-manifolds
and AKSZ-like sigma models the appropriate notion
is that of $Q$-bundles~\cite{Kotov:2007nr}. A bundle 
$\gMfd{E}$ over $\gMfd{X}$ is a $Q$-bundle if both the base
and the total space are $Q$-manifolds and the projection
respects the $Q$-structures:
\begin{equation}
    \big(\gMfd{E}, \Q\big)
    \overset{\pi}{\longrightarrow}
    \big(\gMfd{X}, d_\gMfd{X}\big)
    \InEq{with}
    \pi^* \circ d_\gMfd{X} = \Q \circ \pi^*\,.
\end{equation}
A section $\sigma$ is a $Q$-section if $\sigma^* \circ \Q=d_\gMfd{X}\circ \sigma^*$. In the case where $\gMfd{X}=T[1]\Base$ and the bundle is $\mathbb{Z}$-graded and possibly infinite-dimensional, such an object is known as gauge PDE~\cite{Grigoriev:2019ojp}, see also~\cite{Barnich:2010sw,Barnich:2004cr}, and is known to proved a very flexible version of the Batalin--Vilkovisky formalism at the level of equations of motion. In particular, $Q$-sections are solutions of the gauge PDE while vector fields of the form $[Q,Y]$ generate their gauge transformations.

The sigma-model setup described above corresponds to a trivial $Q$-bundle,
\begin{equation}
    \pi:\gMfd{E} \to \gMfd{X}\,, 
    \qquad     
    \gMfd{E} = \gMfd{M} \times \gMfd{X}\,,
    \qquad 
    \Q = d_\gMfd{X} + Q\,,
\end{equation}
If $\big(\gMfd{M},Q\big)$ is in addition equipped with a closed and $Q$-invariant form $\Omega$, it can be considered as a form on $\gMfd{E}$ by taking its pullback by the canonical projection. Although we limit ourselves to the sigma-model setup most of the considerations  are valid for a generic $Q$-bundle equipped with a closed and $Q$-invariant form $\Omega$ of total degree $n+1$. In particular, the properties of Chern--Weil map remain intact if one replaces the target with the total space and restricts maps to be sections so that ~\eqref{eq:action_multi-AKSZ} is well-defined for a general $Q$-bundle. Of course, in the trivial case this is just reformulation of the same action. Note that when restricted to sections, the Chern--Weil map vanishes on differential forms pulled back from the base. More precisely, for any $\alpha \in \Forms^k,k>0$ one has $\sigma^*_W(\pi^*\alpha)=0$. Indeed,
\begin{equation}
    \sigma^*_W(\pi^*\alpha)
    = p\big(e^{\imath_{d_\gMfd{X}}}\sigma^* 
        e^{-\imath_\Q} \pi^*\alpha\big)
    = p\big(e^{\imath_{d_{\gMfd{X}}}}\sigma^* 
        \pi^* e^{-\imath_{d_\gMfd{X}}} \alpha\big)
    =p(\alpha) = 0\,,
\end{equation}
where we used the fact that $\sigma$ is a section of $\pi$
(and hence $\sigma^* \circ \pi^*
= (\pi \circ \sigma)^*=\1_{\Forms\gMfd{X}}$), and that $\alpha$
is of non-zero form degree by assumption, which implies
that its projection under $p$ vanishes. To show this, let us first note that since $\Q$ projects
onto $d_\gMfd{X}$ under $\pi$, we have the identity
\begin{equation}
    \imath_\Q\pi^* = \pi^*\imath_{d_\gMfd{X}}
    \Implies e^{-\imath_\Q} \pi^*
    = \pi^*e^{-\imath_{d_\gMfd{X}}}\,.
\end{equation}
It follows, $\sigma^*_W(\cI)=0$, where $\cI$ is the ideal in $\Forms^\bullet\cE$ generated by forms from the base and hence is well-defined on vertical forms defined as the following quotient: $\Forms^\bullet\cE/\cI$.

For $\Omega=\omega\in\Forms^2\gMfd{M}$,
the action \eqref{eq:action_multi-AKSZ}
reduces to the standard AKSZ action \eqref{eq:AKSZ}, or its presymplectic version~\cite{Alkalaev:2013hta,Grigoriev:2020xec},
recalled in the Introduction. In this case, $\Theta$ has a $1$-form component $\Theta_{\smbr{1}}=\chi$, and a $0$-form component $\Theta_{\smbr{0}}=\mathcal{L}$, which verify
\begin{equation}
    \dR\chi = \omega\,,
    \qquad 
    \dR\mathcal{L} + \imath_Q\omega = 0\,,
    \qquad 
    \imath_Q\imath_Q\omega = 0\,,
\end{equation}
so that the last relation above is nothing but the condition
that the Hamiltonian $\mathcal{L}$ for the homological
vector field $Q$ self-commutes with respect to the Poisson bracket 
induced by the symplectic form $\omega$. 
In the case of not necessarily trivial $Q$-bundle this action
is still well-defined and is associated with a gauge PDEs
equipped with a presymplectic structure,
see~\cite{Grigoriev:2022zlq, Dneprov:2024cvt} for more details.

\paragraph{Equations of motion and gauge symmetries.}
One can compute the Euler--Lagrange equations resulting
from the action \eqref{eq:action_multi-AKSZ} by considering
variations of the section $\sigma$ induced by vertical
vector fields of degree $0$,
\begin{equation}
    \delta\sigma^* = \sigma^* \circ L_V\,,
    \qquad V \in \Gamma(\mathrm{V}\gMfd{E)}\,,
    \qquad \deg(V)=0\,.
\end{equation}
Using Cartan's homotopy formula, one finds
\begin{equation}\label{eq:interm_variation}
    L_V\Theta = \big(\dR\imath_V + \imath_V\dR\big)\Theta
    = \dR(\cdots) + e^{-\imath_\Q} \imath_V\Omega\,,
\end{equation}
where we used \eqref{eq:exact_form} in the last step.
Consequently, the variation of the action
\begin{equation}
    \delta S = \int_\gMfd{X} p\big(e^{\imath_{d_\gMfd{X}}}
    \sigma^* e^{-\imath_\Q} \imath_V\Omega\big)
    + d_\gMfd{X}(\cdots)
    \equiv \int_\gMfd{X} \sigma^*_W\big(\imath_V\Omega\big)\,,
\end{equation}
where we have discarded the second, $d_\gMfd{X}$-exact, term
as we have assumed the integration over $\gMfd{X}$ 
to be $d_\gMfd{X}$-invariant, and re-wrote the first piece
in terms of the Chern--Weil morphism. One therefore finds
that the equations for motions are given by
\begin{equation}
    \delta S \approx 0 \Iff \Omega(-,F,\dots,F) \approx 0\,,
\end{equation}
where $F$ are the field strengths of the gauge fields
defined by $\sigma^*$. In plain words, the equations of motion
of the action \eqref{eq:action_multi-AKSZ} take the form
of the (pullback of the) multi-presymplectic form $\Omega$
whose `form-legs' are, all but one, contracted with the field
strengths. In components, and assuming for simplicity
that $\Omega$ has only a $p$-form component, these equations read
\begin{equation}
    \Omega_{J I_1 \dots I_{p-1}}(A)\,F^{I_1} \dots F^{I_{p-1}} \approx 0\,.
\end{equation}

Now let us turn our attention to the symmetries 
of the action \eqref{eq:action_multi-AKSZ}. Our claim
is that it is invariant under the gauge transformations 
\begin{equation}\label{eq:gauge_transfo}
    \delta_Y \sigma^* = \sigma^* \circ L_{[\Q,Y]}\,,
    \qquad 
    Y \in \Gamma(\mathrm{V}\gMfd{E})\,,
    \quad \deg(Y) = -1\,,
\end{equation}
generated by vertical vector fields of homological degree $-1$.
To compute the variation of the action
under these gauge transformations, we can pick-up
our previous computation \eqref{eq:interm_variation}
with $V=[\Q,Y]$ and use the identity
\begin{equation}
    e^{\imath_\Q} L_Y e^{-\imath_\Q}
    = L_Y + \imath_{[\Q,Y]}
    \Iff [L_Y, e^{-\imath_\Q}]
    = \imath_{[\Q,Y]} e^{-\imath_\Q}\,,
\end{equation}
to re-write the second term in this equation as
\begin{align}
    \imath_{[\Q,Y]} e^{-\imath_\Q}\Omega
    & = L_Y e^{-\imath_\Q} \Omega
    - e^{-\imath_\Q} L_Y \Omega \\
    & = L_Y \dR \Theta
    - e^{-\imath_\Q} L_Y \Omega
    = \dR(\cdots) - e^{-\imath_\Q}
    L_Y \Omega\,,
\end{align}
so that the variation of the action reads
\begin{equation}
    \delta_Y S[\sigma]
    = \int_{\gMfd{X}} p\big(e^{\imath_{d_\gMfd{X}}}
    \sigma^*e^{-\imath_\Q} L_Y\Omega\big)
    \equiv \int_{\gMfd{X}} \sigma^*_W(L_Y\Omega)\,,
\end{equation}
where again we discarded the $d_\gMfd{X}$-exact terms.
Now assuming that
\begin{equation}
    L_Y\Omega \in \mathcal{I}
    \InEq{where} \mathcal{I}
    = \pi^*\Big(\Forms^{>0}\gMfd{X}\Big)\otimes\Forms^\bullet\gMfd{E}\,,
\end{equation}
is the ideal generated by horizontal forms in the image
of the pullback by the projection map
$\pi:\gMfd{E} \twoheadrightarrow \gMfd{X}$,
of strictly positive form degree, the above variation vanishes.

We have seen above that the Chern--Weil map determined by  a section $\sigma$
vanishes on $\cI$.
As a consequence,
$\sigma^*_W(L_Y\Omega)=0$ and hence the action 
\eqref{eq:action_multi-AKSZ} is invariant under the gauge
transformations \eqref{eq:gauge_transfo} as claimed.

Let us also mention that, if the $(\dR+L_Q)$-closed form $\Omega$
is degenerate---meaning if it has a non-trivial kernel%
---then the action is invariant under
\begin{equation}
    \delta_V\sigma^* = \sigma^* \circ L_V\,,
    \qquad 
    V \in \Ker(\Omega)\,,
    \qquad 
    \deg(V)=0\,,
\end{equation}
which correspond to algebraic / shift symmetries.

\paragraph{Wess--Zumino representation.}
As noticed in \cite[Th. 4.4]{Kotov:2007nr} in the case
of `usual' symplectic AKSZ sigma models, the above action
can also be written as Wess--Zumino term: suppose the source 
manifold is $\gMfd{X}=T[1]\Base$, with
$\Base = \partial\mathscr{Y} \hookrightarrow \mathscr{Y}$,
a $n$-dimensional manifold which is identified as the boundary
of another manifold $\mathscr{Y}$. The above inclusion
defines a morphism of $Q$-manifold between the corresponding
shifted tangent bundles,
\begin{equation}
    \begin{tikzcd}[row sep=large]
        \big(T[1]\Base, d_\Base\big) \ar[r, hook, "i"]
        & \big(T[1]\mathscr{Y}, d_\mathscr{Y}\big)\,,
    \end{tikzcd}
    \qquad 
    i^* \circ d_\mathscr{Y} = d_\Base \circ i^*\,.
\end{equation}
Suppose also that there exists an extension
$\bar\sigma:T[1]\mathscr{Y} \longrightarrow \gMfd{E}$
of a section $\sigma:T[1]\Base \longrightarrow \gMfd{E}$
factorising the latter through the above inclusion, i.e.
\begin{equation}
    \begin{tikzcd}[row sep=large]
        T[1]\Base \ar[dr, swap, "\sigma"]
        \ar[r, hook, "i"] & T[1]\mathscr{Y}
        \ar[d, "\bar \sigma"] \\ & \gMfd{E} 
    \end{tikzcd}
\end{equation}
then the action \eqref{eq:action_multi-AKSZ} can be expressed
as the integral over $T[1]\mathscr{Y}$ of the pullback
by the Chern--Weil morphism of the form $\Omega$. Indeed,
using this factorisation and the fact that $i$ is a $Q$-morphism,
one finds
\begin{equation}
    p_\Base \circ e^{\imath_{d_\Base}} \circ \sigma^*
    = p_\Base \circ e^{\imath_{d_\Base}} \circ i^* \circ\bar\sigma^*
    = p_\Base \circ i^* \circ e^{\imath_{d_\mathscr{Y}}} \circ\bar\sigma^*
    = i^* \circ p_\mathscr{Y} \circ e^{\imath_{d_\mathscr{Y}}} \circ\bar\sigma^*
\end{equation}
where we have denoted the projection of forms on $T[1]\Base$
and $T[1]\mathscr{Y}$ onto their $0$-form components
as $p_\Base$ and $p_\mathscr{Y}$ respectively. Next, 
we can use Stokes' theorem
\begin{equation}
    S[\sigma] = \int_{T[1]\Base} i^* \circ p_\mathscr{Y} 
    \big(e^{\imath_{d_\mathscr{Y}}} \bar\sigma^* \Theta\big)
    = \int_{T[1]\mathscr{Y}} d_\mathscr{Y}\,p_\mathscr{Y} 
    \big(e^{\imath_{d_\mathscr{Y}}} \bar\sigma^* \Theta\big)
    = \int_{T[1]\mathscr{Y}} p_\mathscr{Y} 
    \big(e^{\imath_{d_\mathscr{Y}}} \bar\sigma^* \dR\Theta\big)
\end{equation}
where we used \eqref{eq:exp} and \eqref{eq:proj}
in the last step. Using the defining relation \eqref{eq:exact_form}
of $\Theta$, we therefore end up with
\begin{equation}
    S[\sigma] = \int_{T[1]\Base} 
        p\big(e^{\imath_{d_\Base}}\sigma^*\Theta\big)
    = \int_{T[1]\mathscr{Y}} \bar\sigma_W^*(\Omega)\,,
\end{equation}
which is nothing but the multisymplectic counterpart
of Th. 4.4 of~\cite{Kotov:2007nr}. The analogous representation is well-known for generic multi-symplectic action, see e.g.~\cite{Gotay:1997eg}, and multisymplectic actions associated to PDEs equipped with symplectic currents, see Section~\ref{sec:PDE}.

\section{Examples}
\label{sec:examples}

\subsection{Higher-dimensional Chern--Simons}
A first, simple yet important, class of examples
of multisymplectic AKSZ sigma models is that
of higher-dimensional Chern--Simons theories.
For instance, in $n=5$ dimensions the action takes the form%
\footnote{This action was studied in the context of gravity
in, for instance, \cite{Chamseddine:1989nu, Chamseddine:1990gk}.}
\begin{equation}\label{eq:CS5d}
    S_{\smf CS_{5d}}[A]
    = \int_\Base \tfrac16\,\Omega\big(A,\dR A, \dR A\big)
    + \tfrac18\,\Omega\big(A, \dR A, [A,A]_\g\big)
    + \tfrac1{40}\,\Omega\big(A,[A,A]_\g,[A,A]_\g\big)\,,
\end{equation}
where $\Omega \in (S^3\g^*)^\g$ is a $\g$-invariant polynomial
homogeneous of degree $3$. Equivalently,
$\Omega$ is a trilinear form on $\g$ which is both symmetric
and ad-invariant, i.e.
\begin{equation}
    \Omega\big([x,u]_\g,v,w\big) + \Omega\big(u,[x,v]_\g,w\big)
    + \Omega\big(u,v,[x,w]_\g\big) = 0\,,
    \qquad 
    \forall\, x, u, v, w \in \g\,,
\end{equation}
or in components
\begin{equation}
    f_{d(a}{}^e\,\Omega_{bc)e} = 0\,.
\end{equation}
This invariance property of $\Omega$ ensures
that the action \eqref{eq:CS5d} is invariant
under the usual gauge transformations
\begin{equation}\label{eq:gauge_CS}
    \delta_\varepsilon A = \dR \varepsilon + [A,\varepsilon]_\g\,,
    \qquad
    \varepsilon \in \Functions(\Base) \otimes \g\,,
\end{equation}
\emph{up to} a boundary term---exactly like in the $3d$ case.

Let us try to make sense of the various terms from the point
of view of the target space $\g[1]$. As mentioned 
in the introduction, the invariant polynomial $\Omega$ 
defines a \emph{closed} $3$-form on $\g[1]$, preserved by
the Lie derivative along the homological vector field 
$Q$ given in \eqref{eq:g[1]},
\begin{equation}
    \Forms^3 \g[1] \ni \Omega_{\smbr{3}}
    = \tfrac16\,\Omega_{abc}\dR c^a \dR c^b \dR c^c\,,
    \qquad 
    \dR\Omega_{\smbr{3}} = 0 = L_Q\Omega_{\smbr{3}}\,.
\end{equation}
The first, two-derivative, term of the action \eqref{eq:CS5d}
clearly originates from a `potential'
$\Theta_{\smbr{2}} \ni \Forms^2 \g[1]$ for the $3$-form, 
meaning 
\begin{equation}
    \Omega_{\smbr{3}} = \dR\Theta_{\smbr{2}}\,,
    \InEq{i.e.}
    \Theta_{\smbr{2}} = \tfrac16\,\Omega_{abc}\,
    c^a \dR c^b \dR c^c\,.
\end{equation}
A simple computation shows that the second term
of the action \eqref{eq:CS5d}, comes from a Hamiltonian
$\Theta_{\smbr{1}} \in \Forms^1 \g[1]$ for $Q$
with respect to the $3$-form $\Omega_{\smbr{3}}$,
\begin{equation}
    \Theta_{\smbr{1}} = \tfrac18\,\Omega_{abc}\,
    c^a \dR c^b [c,c]_\g^c\,,
    \qquad
    [c,c]_\g^a := f_{bc}{}^a\,c^b c^c
    \Implies
    \imath_Q\Omega_{\smbr{3}}
        + \dR\Theta_{\smbr{1}} = 0\,,
\end{equation}
as a consequence of the $\g$-invariance of $\Omega_{\smbr{3}}$.
So far, this is simply the higher-form generalisation
of the data needed to define an AKSZ sigma model, namely
a potential for the symplectic form, and a Hamiltonian
for the homological vector field, the only difference being
that the latter is now itself a $1$-form in the target space.
Finally, let us turn our attention to the third and last piece
of the action \eqref{eq:CS5d}. It involves \emph{two} insertions
of the Lie bracket in the trilinear form $\Omega$,
or equivalently \emph{two} contractions of $Q$
with $\Omega_{\smbr{3}}$. Here again, a simple computation
shows that
\begin{equation}\label{eq:chi0-CS5d}
    \Theta_{\smbr{0}} = \tfrac{1}{40}\,\Omega_{abc}\,
    c^a [c,c]_\g^b [c,c]_\g^c
    \Implies
    \tfrac12\imath_Q\imath_Q \Omega_{\smbr{3}}
    - \dR\Theta_{\smbr{0}} = 0\,,
\end{equation}
as a consequence of the $\g$-invariance of $\Omega$,
and hence the Chern--Simons action \eqref{eq:CS5d}
takes the form
\begin{equation}\label{eq:CS5d_AKSZ}
    S_{\smf CS_{5d}}[\sigma]
    = \int_{T[1]\Base} \Big(\tfrac12\imath_{d_\Base} 
        \imath_{d_\Base} \sigma^*\Theta_{\smbr{2}}
    + \imath_{d_\Base}\sigma^*\Theta_{\smbr{1}}
    + \sigma^*\Theta_{\smbr{0}}\Big)\,,
\end{equation}
where $\sigma: T[1]\Base \longrightarrow \g[1]$
is a degree-preserving morphism of graded manifolds,
completely specified by
\begin{equation}
    \sigma^*(c^a) = \theta^\mu A_\mu^a(x)\,,
\end{equation}
with $(x^\mu,\theta^\mu)$ being coordinates of degree $0$ and $1$
respectively. In other words, the components $A_\mu^a(x)$
are indeed those of a $\g$-valued $1$-form on $\Base$ as expected.
In summary, the $5d$ Chern--Simons action can be obtained
from the multisymplectic $Q$-manifold
$\big(\g[1], Q, \Omega_{\smbr{3}}\big)$,
as an AKSZ-type action given by the pullback
of the $k$-forms $\Theta_{\smbr{k}} \in \Forms^k \g[1]$
for $k=0, 1$ and $2$ obeying
\begin{equation}\label{eq:condition_CS5d}
    \left\{
    \begin{aligned}
        \dR\Theta_{\smbr{2}}
            & = \Omega_{\smbr{3}}\,, \\
        \dR\Theta_{\smbr{1}}
            & = -\imath_Q\Omega_{\smbr{3}}\,, \\
        \dR\Theta_{\smbr{0}}
            & = \tfrac12\imath_Q\imath_Q\Omega_{\smbr{3}}\,, \\
        0\,\,\ & = -\tfrac16\,
            \imath_Q\imath_Q\imath_Q\Omega_{\smbr{3}}\,,
    \end{aligned}
    \right. \Iff
    \dR\big(\Theta_{\smbr{2}} + \Theta_{\smbr{1}}
    + \Theta_{\smbr{0}}\big) = e^{-\imath_Q}\Omega_{\smbr{3}}\,.
\end{equation}
On top of that, the gauge transformations \eqref{eq:gauge_CS}
are reproduced by considering degree $-1$ vertical vector fields
on $T[1]\Base \times \g[1]$. Since the fibre, $\g[1]$ here,
is concentrated in degree $1$, and the base $T[1]\Base$
in degree $0$ and $1$, vertical vector fields are necessarily
along $\partial/\partial c^a$ with components of degree $0$,
and hence
\begin{equation}
    Y = \varepsilon^a(x)\,\tfrac{\partial}{\partial c^a}
    \Implies
    \big[d_\Base+Q,Y\big]
    = \Big(\theta^\mu\partial_\mu\varepsilon^a(x)
        + f_{bc}{}^a\,c^b \varepsilon^c(x)\Big)\,
    \tfrac{\partial}{\partial c^a}\,.
\end{equation}
Accordingly, the gauge transformation of the field 
$\sigma^*(c^a) = \theta^\mu\,A_\mu^a(x)$ is obtained by
applying the above vector field to the coordinate $c^a$
and pulling back the result by $\sigma^*$, leading to 
\begin{equation}
    \delta_Y\sigma^*(c^a) = \sigma^*\big([d_\Base+Q,Y]c^a\big)
    = \theta^\mu\big(\partial_\mu\varepsilon^a
    + f_{bc}{}^a\,A_\mu^b \varepsilon^c\big)\,,
\end{equation}
which is indeed the expected gauge transformation
\eqref{eq:gauge_CS} upon using the isomorphism
between functions on $T[1]\Base$ and forms on $\Base$.

The Chern--Simons functional on a manifold $\Base$
of dimension $\dim\Base=2p-1$ is obtained similarly, 
from an invariant polynomial $\Omega \in (S^p\g^*)^\g$,
i.e. a $p$-linear symmetric form on $\g$ verifying
\begin{equation}
    0 = \sum_{k=1}^p \Omega(x_1, \dots,
        x_{k-1}, [x_k,y]_\g, x_{k+1}, \dots, x_p)\,,
    \qquad 
    \forall x_1,\dots,x_p,y \in \g\,,
\end{equation}
considered as a $\Q$-invariant closed $p$-form 
of homological degree $|\Omega_{\smbr{p}}|_{\g\smbr{1}}=p$,
\begin{equation}
    \Omega_{\smbr{p}} = \tfrac{1}{p!}\,
    \Omega_{a_1 \dots a_p}\,\dR c^{a_1} \dots \dR c^{a_p}\,,
    \InEq{such that}
    \Omega_{c(a_1 \dots a_{p-1}} f_{a_p)b}{}^c = 0\,.
\end{equation}
To recover the associated Chern--Simons form,
all we need to do is to find $\Theta \in \Forms^{<p} \g[1]$
such that $\dR\Theta=e^{-\imath_\Q}\Omega_{\smbr{p}}$.
This can be done relatively easily in this case,
upon using the homotopy operator
\begin{equation}\label{eq:h_CS}
    h(\alpha) = \int_0^1 \tfrac{\dR t}{t}\,
    c^a \tfrac{\partial}{\partial \dR c^a}
    \alpha(t\,c, t\,\dR c)\,,
    \qquad 
    \forall \alpha \in \Forms \g[1]\,,
\end{equation}
for the de Rham differential on $\g[1]$,
leading to
\begin{equation}
    \Theta = h(e^{-\imath_\Q}\Omega_{\smbr{p}})
    = \sum_{k=0}^{p-1} \tfrac{p}{2^k(p+k)}\binom{p-1}{k}
    \Omega\big(c,\underbrace{[c,c]_\g,\dots,[c,c]_\g}_{k},
    \underbrace{\dR c, \dots, \dR c}_{p-k-1}\big)\,.
\end{equation}
One can re-write this expression as follows: 
since $\imath_\Q$ is a (degree $0$) derivation
of $\Forms\g[1]$, its exponential
is an algebra morphism and hence
\begin{equation}
    e^{-\imath_\Q}\Omega_{\smbr{p}} = \tfrac1{p!}\,
    \Omega_{a_1 \dots a_p} f^{a_1}(c) \dots f^{a_p}(c)\,
    \InEq{with}
    f^a(c) := \dR c^a + \tfrac12\,[c,c]^a
    \equiv e^{-\imath_\Q}\dR c^a\,.
\end{equation}
We can then express $\Theta$ as
\begin{equation}
    \Theta = \tfrac1{(p-1)!}\,\int_0^1 \dR t\,
    \Omega_{a_1 a_2 \dots a_p} c^{a_1} f^{a_2}(t\,c)
    \dots f^{a_p}(t\,c)\,,
\end{equation}
which is nothing but the `target space counterpart'
of the standard expression
\begin{equation}
    \mathsf{CS}(A) = p\,\int_0^1 \dR t\,
        \Omega\big(A,F_t[A], \dots, F_t[A]\big)\,,
    \InEq{with}
    F_t[A] := t\,\dR A + \tfrac{t^2}{2}\,[A,A]_\g\,,
\end{equation}
for the Chern--Simons form, similarly as the relation
$\dR\Theta=e^{-\imath_\Q}\Omega_{\smbr{p}}$
is the target space incarnation of the defining relation
$\dR\mathsf{CS}(A) = \Omega\big(F[A],\dots,F[A]\big)$.

Note that one can obtain \emph{holomorphic} Chern--Simons
theories by using the source manifold $\gMfd{X}=T^{0,1}[1]\Base$,
i.e. the shifted \emph{anti-holomorphic} tangent bundle
of an odd-dimensional complex manifold $\Base$,
equipped with a holomorphic volume form
$\mu\in\bigwedge^{2p-1,0}\Base$, where $\dim_\C\Base=2p-1$.
In this case, the algebra of functions on $T^{0,1}[1]\Base$,
is isomorphic to the Dolbeault complex $\bigwedge^{0,\bullet}\Base$,
with the Dolbeault differential inducing a homological vector field
on $T^{0,1}[1]\Base$.

\paragraph{Simple variation and Lovelock gravity.}
Let us consider a simple variation on the previous situation,
by assuming that the Lie algebra $\g$ is a semi-direct sum
\begin{equation}
    \g = \h \inplus \mathfrak{p}\,,
\end{equation}
with $\h$ a Lie subalgebra and $\mathfrak{p}$
and Abelian ideal. Denoting coordinates on $\h[1]$
by $\rho^\alpha$ and on $\mathfrak{p}[1]$ by $\xi^a$, 
we can consider the $p$-form of total degree $n+1$,
\begin{equation}
    \Omega_{\smbr{p}}
    := \Omega_{a_1 \dots a_{n+2-2p}b|\alpha_1 \dots \alpha_{p-1}}\,
    \xi^{a_1} \dots \xi^{a_{n+2-2p}}\,\dR \xi^b\,
    \dR\rho^{\alpha_1} \dots \dR \rho^{\alpha_{p-1}}\,,
\end{equation}
which is closed, as we assume its components to be totally
antisymmetric in its Latin indices, and completely symmetric
in the Greek ones. Invariance under the $Q$-structure
defined by the Chevalley--Eilenberg differential of $\g$
amounts to the fact that these components form a $\g$-invariant
tensor. Now instead of the homotopy operator \eqref{eq:h_CS},
let us consider
\begin{equation}
    h(\varpi) = \int_0^1 \tfrac{\dR t}{t}\,\xi^a\,
    \tfrac{\partial}{\partial\dR\xi^a}\,
    \varpi(t\,\xi,t\,\dR\xi,\rho,\dR\rho)\,,
\end{equation}
which defines a contracting homotopy for the de Rham differential
on $\h[1]$ that we trivially extend to $\g[1]$. This allows us
to find a primitive for $e^{-\imath_Q}\Omega_{\smbr{p}}$, given by
\begin{equation}
    \Theta := h(e^{-\imath_Q}\Omega_{\smbr{p}}) = \tfrac{1}{n-2p}\,
    \Omega_{a_1 \dots a_{n+3-2p}|\alpha_1 \dots \alpha_{p-1}}
    \xi^{a_1} \dots \xi^{a_{n+3-2p}}\,
    f^{\alpha_1}(\rho) \dots f^{\alpha_{p-1}}(\rho)\,.
\end{equation}
In the case of the Poincar\'e algebra,
\begin{equation}
    \g = \iso(n-1,1)\,,
    \qquad 
    \h = \so(n-1,1)\,,
    \qquad 
    \mathfrak{p} = \R^n\,,
\end{equation}
such a $Q$-invariant $p$-form is given by the totally
antisymmetric Levi--Civita symbol
\begin{equation}
    \Omega_{\smbr{p}}
    = \epsilon_{a_1 \dots a_{n+1-2p} b_1 c_1 \dots b_p c_p}\,
    \xi^{a_1} \dots \xi^{a_{n+2-2p}} \dR\xi^{a_{n+3-2p}}
    \dR \rho^{b_1 c_1} \dots \dR\rho^{b_{p-1} c_{p-1}}\,.
\end{equation}
Now for a map $\sigma: T[1]\Base \longrightarrow \iso(d-1,1)[1]$,
which defines a vielbein and a spin-connection via
\begin{equation}
    \sigma^*(\xi^a) = \theta^\mu e_\mu^a(x)\,,
    \qquad 
    \sigma^*(\rho^{ab}) = \theta^\mu \omega_\mu^{ab}(x)\,,
\end{equation}
the functional \eqref{eq:action_multi-AKSZ} associated with
the above $(p-1)$-form $\Theta$ reads
\begin{equation}
    \int_\Base \epsilon_{a_1 \dots a_{n+1-2p} b_1 c_1 \dots b_p c_p}\,
    e^{a_1} \wedge \dots \wedge e^{a_{n+1-2p}} \wedge
        R^{b_1 c_1}[\omega] \wedge \dots \wedge R^{b_p c_p}[\omega]\,,
\end{equation}
where 
\begin{equation}
    R^{ab}[\omega] = \dR \omega^{ab} + \omega^a{}_c \wedge \omega^{cb}\,,
\end{equation}
is the Riemann $2$-form curvature of the spin-connection.
For $p=2$, this is nothing but the Cartan--Weyl / Palatini action
for gravity discussed as a presymplectic AKSZ model
in \cite{Grigoriev:2020xec}, while considering a linear combination
of such functionals with different values of $p$ corresponds
to Lovelock gravity models.

\subsection{MacDowell--Mansouri--Stelle--West action}
\paragraph{Four-dimensional case.}
As a second example of multisymplectic AKSZ sigma model,
let us mention the MacDowell--Mansouri--Stelle--West (MMSW) action
for gravity in $4d$ \cite{MacDowell:1977jt, Stelle:1979va, Stelle:1979aj}. In this case, the relevant target space
$Q$-manifold is 
\begin{equation}
    \gMfd{M}_{\smf MMSW} = \R^5\times \mathfrak{o}(3,2)[1]\,.
\end{equation}
The coordinates are determined by the standard basis
in the orthogonal algebra and its vectorial module,
and are denoted by $V^\AmbInd{A}$
and $\rho^\AmbInd{AB}=-\rho^\AmbInd{BA}$, of degree $0$
and $1$ respectively. The action of $\Q$ is determined by
\begin{equation}
    Q V^\AmbInd{A}
        = -\rho^\AmbInd{A}{}_\AmbInd{B} V^\AmbInd{B}\,,
    \qquad 
    Q \rho^\AmbInd{AB}
        = -\rho^\AmbInd{A}{}_\AmbInd{C} \rho^\AmbInd{CB}\,,
\end{equation}
where the indices $\AmbInd{A}=0',0,1,2,3$ are raised
and lowered with the flat metric $\eta_\AmbInd{AB}$
of signature $(-,-,+,+,+)$.
As a multisymplectic form, we take
\begin{equation}
    \Omega_{\smbr{3}} = \dR\Theta_{\smbr{2}}\,,
    \InEq{with}
    \Theta_{\smbr{2}} = \epsilon_\AmbInd{ABCDE} V^\AmbInd{A} 
    \dR\rho^\AmbInd{BC} \dR\rho^\AmbInd{DE}\,,
\end{equation}
and a straightforward computation leads to
\begin{equation}
    -\imath_Q\Omega_{\smbr{3}} = \dR\Theta_{\smbr{1}}\,,
    \InEq{with}
    \Theta_{\smbr{1}}
        = 2\,\epsilon_\AmbInd{ABCDE}\,V^\AmbInd{A} 
        \rho^\AmbInd{B}{}_\AmbInd{F} \rho^\AmbInd{FC} 
        \dR\rho^\AmbInd{DE}\,,
\end{equation}
and
\begin{equation}
    \tfrac12\imath_Q\imath_Q\Omega_{\smbr{3}}
        = \dR\Theta_{\smbr{0}}\,,
    \InEq{with}
    \Theta_{\smbr{0}} = \epsilon_\AmbInd{ABCDE}\,
    V^\AmbInd{A} \rho^\AmbInd{B}{}_\AmbInd{F} \rho^\AmbInd{FC} 
    \rho^\AmbInd{D}{}_\AmbInd{G} \rho^\AmbInd{GE}
\end{equation}
upon using the invariance of $\epsilon_\AmbInd{ABCDE}$
under $\mathfrak{o}(3,2)$. Since the $3$-form $\Omega_{\smbr{3}}$
has homological degree $2$, it has the correct total degree 
to define an action on a four-dimensional spacetime $\Base$:
with the above data, the formula \eqref{eq:action_multi-AKSZ} yields
\begin{equation}
    S[\omega,V] = \int_\Base \epsilon_\AmbInd{ABCDE}\,
    V^\AmbInd{A}\, R^\AmbInd{BC} \wedge R^\AmbInd{DE}\,,
    \qquad 
    R^\AmbInd{AB} := \dR\omega^\AmbInd{AB}
        + \omega^\AmbInd{A}{}_\AmbInd{C} \wedge \omega^\AmbInd{CB}\,,
\end{equation}
thereby reproducing the MMSW  action
for $4d$ gravity---modulo the subtlety of constraining
the norm of the `compensator field' $V^\AmbInd{A}$. Let us also mention the presymplectic AKSZ formulation~\cite{Alkalaev:2013hta} of the MMSW gravity, which gives an equivalent first order action.

\paragraph{Higher-dimensional generalisation.}
Taking a step back, it seems that the above construction
could be carried out similarly for any Lie algebra $\g$
and a module $\cV$ thereof,
with $\Omega \in (\cV^* \otimes S^2\g^*)^\g$
an invariant symmetric bilinear form on $\g$
with values in $\cV^*$, or equivalently a Chevalley--Eilenberg
$0$-cocycle valued in the representation $\cV^* \otimes S^2\g^*$.
In this situation, 
\begin{equation}
    \gMfd{M} = \cV \times \g[1]\,,
\end{equation}
is a $Q$-manifold with degree-$0$ coordinates $V^I$
for the $\cV$ factor, and degree-$1$ coordinates $c^a$
for the $\g[1]$ part, and whose homological vector field
is given by
\begin{equation}
    Q V^I = -c^a\,r_{a\,J}{}^I\,V^J\,,
    \qquad 
    Q c^a = -\tfrac12\,f_{bc}{}^a\,c^b c^c\,,
\end{equation}
where $r_{a\,I}{}^J$ and $f_{bc}{}^a$ are, respectively,
the components of the representation
$r:\g \longrightarrow \End(\cV)$ and the structure constants
of $\g$, in the basis corresponding to the coordinate system 
introduced above. Let us consider the $p$-form
\begin{equation}
    \Forms^p \gMfd{M} \ni \Omega_{\smbr{p}}
        = \Omega_{I_1 \dots I_{p-k}|a_1 \dots a_k}\,
        \dR V^{I_1} \dots \dR V^{I_{p-k}}
        \dR c^{a_1} \dots \dR c^{a_k}\,,
    \qquad 
    \deg(\Omega_{\smbr{p}}) = k\,,
\end{equation}
of homological degree $k$, which is obviously closed
and exact. Its invariance under $Q$,
\begin{equation}
    L_Q \Omega_{\smbr{p}} = 0 
    \Iff 
    (p-k)\,r_{b\,[I_1|}{}^J
    \Omega_{J|I_2 \dots I_{p-k}]|a_1 \dots a_k}
    + k\,f_{b(a_1|}{}^c\,
    \Omega_{I_1 \dots I_{p-k}|a_2 \dots a_k)c} = 0\,,
\end{equation}
is equivalently to the requirement that $\Omega_{\smbr{p}}$
defines a Chevalley--Eilenberg $0$-cocycle valued in
$\wedge^{p-k} \cV^* \otimes S^k\g^*$, that is
$\Omega_{\smbr{p}} \in (\wedge^{p-k} \cV^* \otimes S^k\g^*)^\g$.
As a potential $(p-1)$-form for $\Omega_{\smbr{p}}$,
let us choose
\begin{equation}
    \Forms^{p-1} \gMfd{M} \ni \Theta_{\smbr{p-1}}
        = V^J\Omega_{JI_1 \dots I_{p-k-1}|a_1 \dots a_k}\,
        \dR V^{I_1} \dots \dR V^{I_{p-k-1}}
        \dR c^{a_1} \dots \dR c^{a_k}\,,
\end{equation}
by analogy with the previous example
of MacDowell--Mansouri--Stelle--West gravity.
In order to find the completion of this potential
to an inhomogeneous form verifying
$\dR\Theta=e^{-\imath_Q}\Omega_{\smbr{p}}$,
we can use the operator
\begin{equation}
    h(\alpha) := \int_0^1 \tfrac{\dR t}{t}\,
    V^I \tfrac{\partial}{\partial\dR V^I}
    \alpha(t\,V, t\,\dR V, c, \dR c)\,,
    \qquad 
    \forall \alpha \in \Forms(\cV \times \g[1])\,,
\end{equation}
which defines a contracting homotopy for the de Rham
differential on $\cV \times \g[1]$, namely it satisfies
$\dR \circ h + h \circ \dR = \1$ on forms
with non-zero components in the direction of $\cV$.
This contracting homotopy obviously reproduces the choice
of potential $\Theta_{\smbr{p-1}}$, and since
\begin{equation}
    e^{-\imath_Q}\Omega_{\smbr{p}}
    = \Omega_{I_1 \dots I_{p-k}|a_1 \dots a_k}
    DV^{I_1} \dots DV^{I_{p-k}} f^{a_1}(c) \dots f^{a_k}(c)\,,
\end{equation}
where
\begin{equation}
    DV^I := \dR V^I + c^a\,r_{a\,J}{}^I V^J\,,
\end{equation}
is homogeneous in its dependency in $V^I$,
the action of the contracting homotopy is particularly simple,
and yields
\begin{equation}\label{eq:primitive_MMSW}
    \Theta = h(e^{-\imath_Q}\Omega_{\smbr{p}})
    = \Omega_{I_1 I_2 \dots I_{p-k}|a_1 \dots a_k}
    V^{I_1} DV^{I_2} \dots DV^{I_{p-k}}
    f^{a_1}(c) \dots f^{a_k}(c)\,.
\end{equation}
Specialising to the case $k=2$ and $p=n-1$, 
the above data gives rise to an action
for a $0$-form $C^I = \sigma^*(V^I)$
and a $1$-form gauge field $A^a = \sigma^*(c^a)$
determined by a map
$\sigma:T[1]\Base \longrightarrow \cV \times \g[1]$
with $\dim\Base=n$, which reads
\begin{subequations}
\begin{align}
    S[C,A] & = \int_\Base \Omega
    \big(C,E,\dots,E|F[A],F[A]\big) \\
    & \equiv \int_\Base \Omega_{I_1 I_2 \dots I_{n-3}|ab}\,
    C^{I_1} E^{I_2} \wedge \dots E^{I_{n-3}} 
    \wedge F^a \wedge F^b\,,
\end{align}
\end{subequations}
where we used the notation
\begin{equation}\label{eq:frame}
    E^I := \dR C^I + A^a\,r_{a\,J}{}^I\,C^J\,,
    \qquad 
    F[A]:=\dR A + \tfrac12[A,A]_\g\,.
\end{equation}
The above action is obviously invariant under 
\begin{equation}
    \delta_{\varepsilon} A = \dR \varepsilon + [A,\varepsilon]_\g\,,
    \qquad 
    \delta_\varepsilon C = r(\varepsilon)C\,,
    \InEq{with}
    \varepsilon \in \Functions(\Base) \otimes \g\,,
\end{equation}
as a consequence of the $\g$-invariance of $\Omega$.
For $\g=\mathfrak{o}(n-1,2)$, $\cV=\R^{n+1}$ and
the $(n-2)$-form 
\begin{equation}
    \Omega_{\smbr{n-2}}
    =\epsilon_\AmbInd{ABCD E_1 \dots E_{n-3}}
    \dR V^\AmbInd{E_1} \dots \dR V^\AmbInd{E_{n-3}}
    \dR \rho^\AmbInd{AB} \dR\rho^\AmbInd{CD}\,,
\end{equation}
the inhomogeneous form of total degree $n$ obtained
by \eqref{eq:primitive_MMSW} reads
\begin{equation}
    \begin{aligned}
        \Theta = \epsilon_\AmbInd{ABCD E_1 \dots E_{n-3}}
        V^\AmbInd{E_1} (\dR V^\AmbInd{E_1}
        + \rho^\AmbInd{E_1}{}_\AmbInd{F_1} V^\AmbInd{F_1})
        \ldots &\,(\dR V^\AmbInd{E_{n-3}}
        + \rho^\AmbInd{E_{n-3}}{}_\AmbInd{F_{n-3}}
        V^\AmbInd{F_{n-3}}) \\
        & \qquad \times (\dR \rho^\AmbInd{AB}
        + \rho^\AmbInd{A}{}_\AmbInd{G}\rho^\AmbInd{GB}) 
        (\dR\rho^\AmbInd{CD}
        + \rho^\AmbInd{C}{}_\AmbInd{H}\rho^\AmbInd{HD})\,,
    \end{aligned}
\end{equation}
and reproduces the higher-dimensional generalisation
of the MacDowell--Mansouri--Stelle--West action
discussed previously once pulled back a map
$\sigma: T[1]\Base \longrightarrow \gMfd{M}$
for a spacetime $\Base$ of dimension $\dim\Base=n$,
i.e.
\begin{equation}
    S[\omega,V] = \int_\Base \epsilon_\AmbInd{ABCDE_1 \dots E_{n-4}}
    V^\AmbInd{E_1} E^\AmbInd{E_2} \wedge
        \dots \wedge E^\AmbInd{E_{n-4}}
        \wedge R^\AmbInd{AB} \wedge R^\AmbInd{CD}\,,
    \qquad 
    E^\AmbInd{A} := \dR V^\AmbInd{A}
        + \omega^\AmbInd{A}{}_\AmbInd{B} V^\AmbInd{B}\,.
\end{equation}
In this situation, we see that the combination $E$
of \eqref{eq:frame} gives rise in this case to the frame
defined by the covariant derivative of the compensator field.

\subsection{Self-dual gravity and its higher spin extension}
Self-dual gravity in four dimensions with the cosmological constant 
\cite{Krasnov:2016emc}, and without \cite{Krasnov:2021cva},
can also be obtained as a multisymplectic AKSZ model.
The relevant $Q$-manifold has a structure similar to that
of MacDowell--Mansouri--Stelle--West formulation of gravity,
namely 
\begin{equation}\label{eq:gMfd_SDGR}
    \gMfd{M}_{\smf SDGR} = \cV_5 \times \g[1]\,,
\end{equation}
where $\g=\su(2)$ for the Euclidean signature
and $\cV_5$ is its dimension $5$ irreducible representation.
Indeed, in Euclidean signature the four-dimensional Lorentz algebra
is replaced with the orthogonal algebra
$\so(4) \cong \su(2) \oplus \su(2)$, so that $\g$ corresponds
to the subspace in which the self-dual component
of the spin-connection takes values (i.e. one of the $\su(2)$
factors) and $\cV_5$ to the irrep in which the self-dual part
of the Weyl tensor takes values.

The graded manifold \eqref{eq:gMfd_SDGR} is endowed with
a homological vector field, whose action on coordinates
$c^{AB}$ and $V^{ABCD}$ with $A,B,\dots=1,2$,
of degree $1$ and $0$ respectively, is given by
\begin{align}
    Q c^{AB} & = -c\fud{A}{C} c^{CB}\,,
    && Q V^{ABCD} = -4\,c\fud{(A}{E} V^{BCD)E}\,.
\end{align}
where the two-component spinor indices are raised and lowered
with the $\su(2)$-invariant tensor $\epsilon^{AB}=-\epsilon^{BA}$
according to $v^A = \epsilon^{AB} v_B$ and $v_A = v^B \epsilon_{BA}$.%
\footnote{Note that, accordingly, we use the convention
$\epsilon^{AC} \epsilon_{BC} = \delta^A_B$ for the inverse
of this tensor.}
The case of vanishing cosmological constant is even simpler,
as we take the same graded manifold but with trivial $Q$-structure,
\begin{align}
    Q c^{AB}&= 0\,, && Q V^{ABCD}=0\,,
\end{align}
which amounts to considering the contraction of $\so(4)$
obtained by making the $\su(2)$ factor accounting for the self-dual
piece of the spin-connection Abelian \cite[Sec. 4]{Krasnov:2021cva}.
This $Q$-manifold is also equipped with an invariant $3$-form,
\begin{equation}
    \Omega = \dR V^{ABCD} \dR c_{AB} \dR c_{CD}\,,
\end{equation}
which verifies
\begin{equation}
    e^{-\imath_Q}\Omega = \dR\Theta\,,
    \InEq{with}
    \Theta = V^{ABCD}\,(\dR c_{AB} + \lambda\,c_{AE}\,c\fud{E}{B})
    (\dR c_{CD} + \lambda\,c_{AF}\,c\fud{F}{D})\,,
\end{equation}
where we introduced the parameter $\lambda$ which we can set to $0$
for vanishing cosmological constant, and $1$ otherwise.
Denoting the fields resulting from the pullback
by $\sigma: T[1]\Base \longrightarrow \gMfd{M}_{\smf SDGR}$
of the coordinate functions
\begin{equation}
    \sigma^*(c^{AB}) = \theta^\mu\omega_\mu^{AB}(x)\,,
    \qquad 
    \sigma^*(V^{ABCD}) = \Psi^{ABCD}(x)\,,
\end{equation}
one finds that the action \eqref{eq:action_multi-AKSZ}
obtained from the above data reads
\begin{equation}
    S[\omega,\Psi] = \int_{\Base} \Psi^{ABCD}\,R_{AB} \wedge R_{CD}\,,
    \qquad 
    R_{AB} := \dR \omega_{AB}
        + \lambda\,\omega_{AC} \wedge \omega_\fud{C}{B}\,,
\end{equation}
that is, we recover the action for the self-dual gravity
with the cosmological constant\cite{Krasnov:2016emc}.
Similarly, the second $Q$-manifold with the same $\Omega$ gives
\begin{equation}
    S[\omega,\Psi] = \int_{\Base} \Psi^{ABCD}\,
    \dR \omega_{AB} \wedge \dR \omega_{AB}\,,
\end{equation}
which is the self-dual gravity with the vanishing cosmological 
constant \cite{Krasnov:2021cva}.
These actions are invariant under the gauge transformations
\begin{equation}
    \delta_\xi\omega_{AB}
        = \dR\xi_{AB} + 2\lambda\,\omega_{C(A}\xi^{C}{}_{B)}\,,
    \qquad 
    \delta_\xi\Psi^{ABCD} = 4\,\xi_{E}{}^{(A}\Psi^{BCD)E}\,,
\end{equation}
which are accounted for by vertical vector fields 
of the form $Y = \xi_{AB}(x)\,\tfrac{\partial}{\partial c_{AB}}$
when used in the formula \eqref{eq:gauge_transfo}.
The action are also invariant under diffeomorphisms,
\begin{equation}\label{eq:diffeo}
    \delta_\eta\omega_{AB} = L_\eta\omega_{AB}\,,
    \qquad 
    \delta_\eta\Psi^{ABCD} = L_\eta\Psi^{ABCD}\,,
    \InEq{with}
    \eta \in \Gamma(T\Base)\,,
\end{equation}
which can be represented in terms of the degree $-1$
vector field on $T[1]\Base$ as,
\begin{equation}
    y = \eta^\mu(x) \tfrac{\partial}{\partial\theta^\mu}
    \Implies
    [y, d_\Base] = \eta^\mu(x) \tfrac{\partial}{\partial x^\mu}
    + \theta^\mu \partial_\mu\eta^\nu(x)
        \tfrac{\partial}{\partial\theta^\nu}\,,
\end{equation}
i.e. the Lie bracket of $y$ with the homological vector field
$d_\Base$ is nothing but the vector field on $T[1]\Base$
corresponding to the Lie derivative along $\eta$
(acting on forms on $\Base$). Transforming the section $\sigma$ 
according to
\begin{equation}
    \delta_y\sigma^* = L_{[y,d_\Base]} \circ\sigma^*
\end{equation}
reproduces the Lie derivative of $\omega_{AB}$
and $\Psi^{ABCD}$ when acting on the coordinates $c_{AB}$
and $V^{ABCD}$ respectively, and hence yields \eqref{eq:diffeo}.

Let us note that the Krasnov formulation of the self-dual gravity with the cosmological constant naturally emerges from the Pleba\'nski action \cite{Plebanski:1977zz}. The latter is straightforwardly another example of a multisymplectic theory.

\paragraph{Higher spin extension.}
Self-dual gravity in four dimensions
admits an higher spin extension \cite{Krasnov:2021nsq},%
\footnote{It was fist discovered in the light-cone gauge
in \cite{Ponomarev:2017nrr} as a truncation of chiral higher-spin 
gravity \cite{Metsaev:1991mt, Metsaev:1991nb, Ponomarev:2016lrm}.}
whose action takes the form
\begin{equation}\label{eq:action_SDHSGR}
    S[\omega, \Psi] = \int_\Base \pmb\langle\Psi, 
    F[\omega] \wedge F[\omega]\pmb\rangle\,,
    \qquad 
    F[\omega] = \dR\omega + \tfrac12\,\{\omega,\omega\}\,,
\end{equation}
where $\Psi$ and $\omega$ are respectively
a $0$- and $1$- form valued in 
\begin{equation}
    \hs_{\mathtt{Gr}} := \C[y^A]\,,
    \qquad A=1,2\,,
\end{equation}
which is a Poisson algebra, with Poisson bracket
\begin{equation}
    \{f,g\} := \epsilon^{AB}\,\tfrac{\partial f}{\partial y^A}
    \tfrac{\partial g}{\partial y^B}\,,
    \qquad 
    f, g \in \hs_{\mathtt{Gr}}\,.
\end{equation}
Note that this algebra admits a subalgebra, namely the subspace
consisting of even polynomials in $y^A$, which allows one to focus
on bosonic fields only.
The final ingredient used in the action \eqref{eq:action_SDHSGR}
is an invariant form $\pmb\langle-,-\pmb\rangle$ defined
on a pair of elements $f,g \in \C[y^A]$
of the form $f = \sum_{m\geq0} \tfrac1{m!}\,f^{A(m)}\,y_{A(m)}$
and $g = \sum_{k\geq0} \tfrac1{k!}\,g_{A(k)} y^{A(k)}$ as
\begin{equation}
    \pmb\langle f, g \pmb\rangle
    := \sum_{n=0}^\infty \tfrac1{n!}\,f^{A(n)}\,g_{A(n)}\,.
\end{equation}
This action is invariant under 
\begin{equation}
    \delta_{\xi,\eta} \omega = \dR\xi + \{\omega,\xi\}
    + \imath_\eta F[\omega]\,,
    \qquad 
    \delta_{\xi,\eta} \Psi = \Psi \pmb{\circ} \xi
    + \imath_\eta\big(\dR\Psi - \Psi \pmb{\circ} \omega\big)\,,
\end{equation}
where
\begin{itemize}
\item The operation
\begin{equation}
    \pmb{\circ} : \hs_{\mathtt{Gr}}^* \otimes \hs_{\mathtt{Gr}}
    \longrightarrow \hs_{\mathtt{Gr}}^*\,,
\end{equation}
defines a right representation of the Poisson bracket 
on the dual module of $\hs_{\mathtt{Gr}}$ by
\begin{equation}\label{eq:right_action}
    \pmb\langle \psi \pmb{\circ} f, g \pmb\rangle
    = \pmb\langle \psi, \{f, g\} \pmb\rangle\,,
    \qquad 
    \psi, f, g \in \hs_{\mathtt{Gr}}\,,
\end{equation}
where we use the fact that $\pmb\langle-,-\pmb\rangle$
is non-degenerate to identify an element
of $\hs_{\mathtt{Gr}}^*$ as $\pmb\langle\psi,-\pmb\rangle$
for some $\psi \in \hs_{\mathtt{Gr}}$.

\item The shift parameter
$\eta \in \Gamma(T\Base) \otimes \hs_{\mathtt{Gr}}$
is a vector field valued in the higher spin algebra, i.e.
\begin{equation}
    \delta_\eta \omega = \eta^\mu\,
    \tfrac{\partial}{\partial(\dR x^\mu)}\,F\,,
    \qquad 
    \eta = \sum_{s\geq0} \tfrac1{n!}\,
    \eta^\mu_{A_1 \dots A_n}(x)\,y^{A_1} \dots y^{A_n}\,
    \tfrac{\partial}{\partial x^\mu}\,.
\end{equation}
\end{itemize}
We have essentially the same structures at hand
than in the previous cases of MacDowell--Mansouri--Stelle--West
or self-dual gravity, namely a Lie algebra together with
a representation and an invariant form on the latter 
tensored with the symmetric tensor product of the former.
More precisely, we should consider here the graded manifold
\begin{equation}
    \gMfd{M}_{\smf HS-SDGR} = \hs_\mathtt{Gr} \times \hs_\mathtt{Gr}[1]\,,
\end{equation}
which is again concentrated in degree $0$ and $1$.
A basis of $\hs_{\mathtt{Gr}}$ is given by (even) monomials
in $y^A$ tensored which leads to
\begin{equation}
    \hs_\mathtt{Gr}
    = {\rm span}_\C\big\{y^{A_1} \dots y^{A_n}
    \mid n \in \N\big\}
    \qquad\leadsto\qquad 
    \left\{
    \begin{aligned}
        V_{A_1 \dots A_n} && \text{in deg. } 0\,,\\
        c_{A_1 \dots A_n} && \text{in deg. } 1\,,
    \end{aligned}
    \right.
\end{equation}
for coordinates on $\gMfd{M}_{\smf HS-SDGR}$.
Their pullback by $\sigma: T[1]\Base
\longrightarrow \gMfd{M}_{\smf HS-SDGR}$
defines the basic fields participating in the higher spin
self-dual gravity action \eqref{eq:action_SDHSGR}, namely
\begin{equation}
    \sigma^*(c^{A_1 \dots A_n})
        = \theta^\mu\omega_\mu^{A_1 \dots A_n}(x)\,,
    \qquad 
    \sigma^*(V^{A_1 \dots A_n}) = \Psi^{A_1 \dots A_n}(x)\,,
\end{equation}
which are nothing but the components of a $1$-form
and a $0$-form, both valued in $\hs_\mathtt{Gr}$.
Now defining
\begin{equation}
    \Omega: \hs_\mathtt{Gr} \otimes S^2 \hs_\mathtt{Gr}
    \longrightarrow \C 
    \InEq{by}
    \Omega(\psi|f,g) := \pmb\langle\psi, f \cdot g \pmb\rangle\,,
    \qquad \psi,f,g\in\hs_\mathtt{Gr}\,,
\end{equation}
and thinking of the first factor as a representation
of the Lie algebra $\big(\hs_\mathtt{Gr},\{-,-\}\big)$ via
\begin{equation}
    r(f) \psi := -\psi \pmb{\circ} f\,,
    \qquad 
    f,\psi \in \hs_\mathtt{Gr}\,,
\end{equation}
one can check that $\Omega$ obeys the invariance property
\begin{equation}\label{eq:inv}
    \Omega\big(r(\varepsilon)\psi|f,g\big)
    + \Omega(\psi|\{\varepsilon,f\},g\big)
    + \Omega(\psi|f,\{\varepsilon,g\}\big) = 0\,,
\end{equation}
as a consequence of the fact that $\{-,-\}$ is a Poisson bracket
(more specifically it satisfies the Leibniz identity),
and the definition \eqref{eq:right_action}. Equivalently,
we present $\Omega$ as a $3$-form on $\gMfd{M}_{\smf HS-SDGR}$, 
whose explicit expression reads
\begin{equation}
    \Omega_{\smbr{3}} = \sum_{m+n\geq0} \tfrac{1}{(m+n)!}\,
    \dR V^{A_1 \dots A_{m+n}} \dR c_{A_1 \dots A_m}
    \dR c_{A_{m+1} \dots A_{m+n}}\,.
\end{equation}
As expected, the $Q$-structure on $\gMfd{M}$ is nothing but
Chevalley--Eilenberg differential associated with $\hs_\mathtt{Gr}$
in the representation $\big(\hs_\mathtt{Gr},r\big)$,
which schematically takes the form
\begin{subequations}
\begin{align}
    Q c_{A_1 \dots A_n} & = -\tfrac12\,
    \sum_{m=0}^n \binom{n}{m}\,c_{B(A_1 \dots A_m}\,
    c^B{}_{A_{m+1} \dots A_n)}\,,\\
    Q V^{A_1 \dots A_n}
    & = -\sum_{m\geq0} \binom{m+n}{n}V^{B_1 \dots B_m (A_1 \dots A_{n-1}}
        c^{A_n)}{}_{B_1 \dots B_m}\,,
\end{align}
\end{subequations}
and one can check that $L_Q\Omega_{\smbr{3}}=0$
is equivalent to the invariance condition \eqref{eq:inv}.

\subsection{Twistor space formulation of self-dual gravity}
Self-dual gravity, as well as its supergravity extension,
admit a twistor formulation \cite{Mason:2007ct} 
(see also \cite{Herfray:2016qvg,Herfray:2018wxe} for an overview).
Focusing on the non-supersymmetric case, the idea
is to work on a $3$-dimensional almost complex manifold
$\mathcal{PT}$, and consider a line subbundle
$L^* \subset T^*\mathcal{PT}$ in the cotangent bundle
of this complex manifold. Locally, this line subbundle
is specified by a $L$-valued $1$-form 
$\tau \in \Forms^1\mathcal{PT} \otimes L$,
and introducing another $1$-form
$b \in \Forms^1 \mathcal{PT} \otimes (L^*)^{\otimes 3}$
valued in the third tensor power of $L^*$, to play
the role of Lagrange multiplier in the following action
\begin{equation}\label{eq:twistor}
    S[b,\tau] = \int_\mathcal{PT} b \wedge
        \tau \wedge \dR\tau \wedge \dR\tau\,,
\end{equation}
thereby enforcing $\tau\wedge(\dR\tau)^{\wedge2} = 0$
as equations of motion. The latter equation implies
that the almost complex structure on $\mathcal{PT}$
is integrable \cite[Th. 3]{Mason:2007ct}. The action
is invariant under 
\begin{equation}\label{eq:rescaling}
    b \to \tfrac{1}{f^3}\,b\,,
    \qquad 
    \tau \to f\,\tau\,,
    \qquad 
    f \in \Functions(\mathcal{PT},\C)\,,
\end{equation}
which merely expressed the fact that the integrand 
of the action \eqref{eq:twistor} is of weight $0$,
in the sense of homogeneity in the line bundle $L$.

In order to reproduce the above action as a multisymplectic
AKSZ sigma model, let us consider the manifold
\begin{equation}
    \gMfd{M}_{\smf tw.-SDGR}
        = \big(L \oplus (L^*)^{\otimes 3}\big)[1]\,,
\end{equation}
which is concentrated in degree $0$ and $1$,
equipped with the \emph{trivial} $Q$-structure.
Denoting by $x^\mu$ the degree $0$ coordinates,
i.e. coordinate on the base manifold $\mathcal{PT}$,
and by $\lambda$ and $\beta$ the degree $1$ coordinates,
i.e. the coordinates along the fibres of $L$ and $L^*$
respectively, the relevant pair of fields originate
from the pullback by a map $\sigma: T[1]\mathcal{PT}
\longrightarrow \gMfd{M}_{\smf tw.-SDGR}$
of the degree $1$ coordinates, namely
\begin{equation}
    \sigma^*(\lambda) = \theta^\mu\tau_\mu(x)\,,
    \qquad 
    \sigma^*(\beta) = \theta^\mu b_\mu(x)\,.
\end{equation}
The closed form on $\gMfd{M}_{\smf tw.-SDGR}$ of interest 
for us reads
\begin{equation}
    \Omega = \big(\dR\beta\,\lambda-\beta\,\dR\lambda\big)
    \wedge \dR\lambda \wedge \dR\lambda \equiv \dR\Theta\,,
    \InEq{with}
    \Theta =\beta \wedge \lambda
        \wedge \dR\lambda \wedge \dR\lambda\,.
\end{equation}
Note that $\tdeg(\Omega)=7$, as is appropriate to construct
an action functional on a \emph{real} six-dimensional manifold,
which is our case here for $\mathcal{PT}$. One can easily see
that the pullback by $\sigma^*\Theta$ yields back the action
\eqref{eq:twistor}. Another straightforward check is that
the degree $0$ vector field
\begin{equation}
    V = \lambda\tfrac{\partial}{\partial\lambda}
    -3\beta\tfrac{\partial}{\partial\beta}
    \Implies
    \imath_V\Omega=0\,,
\end{equation}
sits in the kernel of $\Omega$, and hence generates a symmetry
of the action. This vector field can be thought of 
as an Euler vector field measuring the ``$L$-weight'' in target space,
and unsurprisingly generates the infinitesimal version
of the rescaling symmetry \eqref{eq:rescaling}.

\subsection{Sparling gravity}
In a couple of papers, Sparling proved that Einstein's
vacuum field equations can be reformulated as the closure
of a certain $3$-form, either on the bundle of orthonormal frames \cite{Sparling:2001a}, or on the spin bundle
\cite{Sparling:2001b, Mason:1990}, of the four-dimensional
spacetime manifold. Here we focus on the latter, and show
that the action functional defined by lifting the closure
of the Sparling $3$-form using a Lagrange multiplier
can be recovered as a multisymplectic AKSZ action.

The Sparling $3$-form is given by
\begin{equation}
    \mathfrak{S} := \nabla\pi_{A'} \wedge\nabla\bar\pi_A
    \wedge e^{AA'}\,,
\end{equation}
where $\pi_{A'}$ and $\bar\pi_A$ are primed and unprimed spinors,
$\nabla$ a not necessarily torsion-free connection and $e^{AA'}$
a vierbein. Introducing a Lagrange multiplier $\Psi$,
one can simply consider the following action functional,
\begin{equation}\label{eq:Sparling_gravity}
    S[\Psi,\pi,\bar\pi,\omega,e]
    = \int_\Base \Psi \wedge \nabla\mathfrak{S}\,,
\end{equation}
whose equations of motion with respect to $\Psi$
impose that $\mathfrak{S}$ be closed, and hence that 
Einstein's equations be satisfied.

In order to reproduce this action as a multisymplectic
AKSZ action, let us introduce the graded manifold
\begin{equation}\label{eq:target_Sparling}
    \gMfd{M}_{\smf Sparling} = \g[1] \times \mathcal{S}
        \times \mathcal{S}' \times \R\,,
\end{equation}
where $\g=\so(4)\cong\su(2)\oplus\su(2)$ in Euclidean signature,
or the complexification of $\sl(2,\C)$ if one avoids fixing
reality conditions, and $\mathcal{S}'$ and $\mathcal{S}$
are the primed and unprimed spinors representations.
The $Q$-structure is given by the Chevalley-Eilenberg differential
of $\g$, i.e. it reads
\begin{equation}
    Q\rho_{AB} = -\rho_{AC}\rho_{B}{}^C\,,
    \qquad 
    Q\rho_{A'B'} = - \rho_{A'C'}\rho_{B'}{}^{C'}\,,
    \qquad 
    Q\xi^{AA'} = -\rho^A{}_B\xi^{BA'} - \rho^{A'}{}_{B'}\xi^{AB'}\,,
\end{equation}
on the degree $1$ coordinates $\rho_{AB}$, $\rho_{A'B'}$
and $\xi_{AA'}$ (with $A,A',\dots=1,2$), and
\begin{equation}
    QV = 0\,,
    \qquad 
    Q\psi_{A'} = -\rho_{A'B'}\,\psi^{B'}\,,
    \qquad 
    Q\bar\psi_A = - \rho_{AB}\,\bar\psi^{B}\,,
\end{equation}
on the degree $0$ coordinate $V$ for the $\R$-factor,
and $\psi_{A'}$ and $\bar\psi_A$ for the $\mathcal{S}'$
and $\mathcal{S}$ factor.
In this case, the relevant form is inhomogeneous 
in form degree: it has a $3$-form and a $4$-form component,
\begin{equation}
    \Omega = \Omega_{\smbr{4}} + \Omega_{\smbr{3}}\,,
    \qquad 
    \tdeg(\Omega) = 5\,,
\end{equation}
respectively given by
\begin{equation}
    \Omega_{\smbr{4}} = \dR V \dR \psi_{A'}
        \dR\bar\psi_A \dR\xi^{AA'}\,,
    \qquad 
    \Omega_{\smbr{3}}
    = \dR V\big(\dR\bar\psi_A \dR\rho_{A'B'}\psi^{B'} 
    - \dR \psi_{A'}\dR\rho_{AB}\bar\psi^B\big)\xi^{AA'}\,.
\end{equation}
One can verify that this form is $(\dR+L_Q)$-closed,
\begin{equation}
    \big(\dR+L_Q)\Omega = 0 
    \Iff 
    \left\{
    \begin{aligned}
        0 & = \dR\Omega_{\smbr{4}}\,,\\
        0 & = \dR\Omega_{\smbr{3}} + L_Q\Omega_{\smbr{4}}\,,\\
        0 & = L_Q\Omega_{\smbr{3}}\,,
    \end{aligned}
    \right.
\end{equation}
and that the form of total degree $4$ given by
\begin{equation}
    \Theta = V\,\dR\vartheta\,,
    \InEq{with}
    \vartheta = \big(\dR\psi_{A'} + \rho_{A'B'}\psi^{B'})
    (\dR\bar\psi_A + \rho_{AB}\bar\psi^B)\xi^{AA'}\,,
\end{equation}
which is the target space counterpart of the Sparling $3$-form.
Finally, a map $\sigma: T[1]\Base
\longrightarrow \gMfd{M}_{\smf Sparling}$
from the shifted tangent bundle of the four-dimensional
spacetime manifold $\Base$ to the target space defined
in \eqref{eq:target_Sparling} gives to a vierbein
and a spin-connection, via
\begin{equation}
    \sigma^*(\xi^{AA'}) = \theta^\mu e_\mu^{AA'}(x)\,,
    \qquad 
    \sigma^*(\rho^{AB}) = \theta^\mu \omega_\mu^{AB}(x)\,,
    \qquad 
    \sigma^*(\rho^{A'B'}) = \theta^\mu \omega_\mu^{A'B'}(x)\,,
\end{equation}
as well as the spinors and Lagrange multiplier appearing
in the action \eqref{eq:Sparling_gravity}, via
\begin{equation}
    \sigma^*(V) = \Psi(x)\,,
    \qquad 
    \sigma^*(\psi_{A'}) = \pi_{A'}(x)\,,
    \qquad 
    \sigma^*(\bar\psi_A) = \bar\pi_A(x)\,.
\end{equation}
One can check that applying the construction of the multisymplectic
AKSZ action \eqref{eq:action_multi-AKSZ} with the above data
yields the Sparling gravity action \eqref{eq:Sparling_gravity}.

\section{Relation to multisymplectic actions}
\label{sec:PDE}
In the preceding Sections we have restricted ourselves
to the case where the underlying $Q$-bundle is globally trivial
as a $Q$-bundle. 
In plain terms this means that the target space and the source
are completely detached, i.e. the target space structures
do not depend on the source space coordinates. Although
the generalisation to the case of generally nontrivial $Q$-bundles
is relatively straightforward and gives a multisymplectic extension 
of the presymplectic BV--AKSZ formalism 
of~\cite{Dneprov:2024cvt,Grigoriev:2022zlq} we refrain 
from discussing it in full generality here because it would have 
required introducing additional techniques. Nevertheless, 
it is instructive to sketch the generalisation in the particular 
but very important case where the fibre of the underlying bundle 
is a usual (not graded) manifold or, in other words, the grading 
is horizontal. In this case, the underlying geometry 
is that of a bundle equipped with a flat connection,
i.e. PDE defined geometrically. In particular, the construction
can be entirely reformulated in terms of the conventional 
differential geometry without resorting to supergeometry. 
Moreover, it turns out that in this case the multisymplectic
AKSZ actions discussed in this work are quite similar 
to multisymplectic actions known in the context 
of multisymplectic formalism and geometry
of PDEs~\cite{{Gotay:1997eg,Bridges:2010,Roman-Roy:2005vwe}}.
In the standard symplectic case, i.e. $p=2$, this relation
was observed already  in~\cite{Grigoriev:2016wmk}, 
see also~\cite{Grigoriev:2021wgw,Dneprov:2024cvt}.

Let us first recall what is usually meant by multi-symplectic 
systems. Let ${\tilde \cE} \longrightarrow \Base$ be a fibre bundle 
equipped with the $n=\dim(\Base)$-form $\tilde\Theta$ such that
it vanishes on any $p$ vertical vectors and such that
$\tilde\Omega\equiv \dR\tilde\Theta$ is nondegenerate
and vanishes on any $p+1$ vertical vectors. It follows
that this data defines a natural functional on the space
of sections of ${\tilde \cE}$:
\begin{equation}\label{ms-action}
    S[\tilde\sigma] = \int_\Base \tilde\sigma^*(\tilde\Theta)\,, 
    \qquad \tilde\sigma: \Base \longrightarrow {\tilde \cE}
\end{equation}
Of course, this action is perfectly defined for any $n$-form
$\tilde\Theta$ but under the above condition the integrand
is of homogeneity $p-1$ in $\dR z^I(x)$,
$z^I(x) \equiv \tilde\sigma^*(z^I)$,
where $z^I$ are fibre coordinates. The Euler--Lagrange equations
for this action can be formulated as follows:
a section $\tilde\sigma$ is a solution if 
\begin{equation}
    \tilde\sigma^*(i_V\tilde\Omega)=0\,,
\end{equation}
for any vertical vector field $V$ on ${\tilde \cE}$. In particular, 
if a vertical vector field $W$ belongs to the kernel
of $\tilde\Omega$ then $\delta \tilde\sigma^*
=\tilde\sigma^* \circ \epsilon W$ is an algebraic gauge symmetry
of the above action for an arbitrary function
$\epsilon\in \Functions({\tilde \cE})$. Moreover,
if $W$ is a regular vector field, then the action is effectively 
defined on the quotient of ${\tilde \cE}$ by the integral curves
of $W$, at least locally. This explains why it is natural
to require $\tilde\Omega$ to be non-degenerate in the first place.

In applications to field theory, bundle ${\tilde \cE}$ 
usually comes equipped with additional structures.
Let us concentrate on the case when ${\tilde \cE}$
is equipped with a flat connection. In particular,
${\tilde \cE}$ can be naturally interpreted as PDE
and, reciprocally, any regular PDE gives rise
to such a bundle~\cite{Vinogradov:1981,Bocharov:1999}.
Thanks to the flat connection, the differential forms
on ${\tilde \cE}$ are bigraded by the horizontal,
and vertical form degree and the de Rham differential
decomposes as:
\begin{equation}
    \dR=\dh+\dv\,.
\end{equation}

Assume for simplicity that $\tilde\Omega$ is of definite bidegree
and hence is a $(n+1-p,p)$ form satisfying
$\dv\tilde\Omega=0=\dh\tilde\Omega$. It follows
\begin{equation}
    \tilde\Theta=\tilde\Theta_{\smbr{p-1}}
    +\tilde\Theta_{\smbr{p-2}}+\ldots+\tilde\Theta_{\smbr{0}}\,,
\end{equation}
and $\tilde\Omega=\dv \tilde\Theta_{\smbr{p-1}}$.
Then the equations of motion can be written explicitly as
\begin{equation}
    \tilde\sigma^*(\tilde\Omega_{IJ_2\ldots J_p})\,
        E_{\tilde\sigma}^{J_2}\ldots E_{\tilde\sigma}^{J_p}=0\,,
    \qquad
    E_{\tilde\sigma}^I \equiv \tilde\sigma^*\dv z^I
    = \dR\tilde\sigma^*z^I - \tilde\sigma^*\dh z^I\,,
\end{equation}
where we introduced ``vertical coefficients'' of $\tilde\Omega$
as $\tilde\Omega=\tilde\Omega_{I_1 \ldots I_{p}}
\dv z^{I_1}\ldots \dv z^I_{p}$. Note that these coefficients 
are themselves horizontal forms, i.e. $(n+1-p,0)$-forms.
It is easy to see that if $\tilde\sigma$ is such that 
$E_{\tilde\sigma}^I=0$ then $\tilde\sigma$ is a covariantly 
constant section of ${\tilde \cE}$. In particular,
if we interpret ${\tilde \cE}$ as a PDE then
such a $\tilde\sigma$ is a solution. It follows
that the Euler--Lagrange equations for $S[\tilde\sigma]$
are consequences of the equations of motion of ${\tilde \cE}$
seen as a PDE.

The above multisymplectic setup can be easily reformulated
in the supergeometrical terms of $Q$-bundles. Namely,
let $(\cE,Q) \to (T[1]\Base, \dx)$ be a $Q$-bundle such that the grading is horizontal, i.e. the only nonvanishing degree coordinates are the fibre coordinates $\theta^\mu$ of $T[1]\Base$, then the $Q$-structure determines a flat connection in the bundle ${\tilde \cE} \longrightarrow \Base$ which is a restriction of $\cE$ to the zero section of $T[1]\Base$. Note that in our setup ${\tilde \cE} \longrightarrow \Base$ is a bundle of usual (not graded) manifolds. Other way around, given a bundle ${\tilde \cE} \longrightarrow \Base$ equipped with a flat connection, there is a canonically associated $Q$-bundle which can be obtained by pulling back ${\tilde \cE}$ by the canonical projection $T[1]\Base \longrightarrow \Base$ and reinterpreting the connection as the $Q$-structure.

Moreover, in this setup there is a one-to-one relation between differential forms on ${\tilde \cE}$ and vertical forms on $\cE$. More precisely, let $\cI$ denote the ideal of differential forms on $\cE$ generated by the positive degree forms pulled back from the base $T[1]\Base$. The quotient space $\bigwedge^\bullet(\cE)/\cI$ is the space of vertical forms. It is easy to check that it inherits the form degree and the ghost degree. In addition, the de Rham differential is well defined on equivalence classes and makes vertical forms into differential graded commutative algebra. Furthermore, vertical forms on $\cE$ defined in this way are naturally isomorphic to the differential forms on ${\tilde \cE}$. Under this isomorphism the de Rham differential goes to vertical differential on ${\tilde \cE}$, ghost degree goes into horizontal form degree, and $L_Q$
is mapped to the horizontal differential $\dh$, see~\cite{Dneprov:2024cvt} for further details. If $x^\mu, \theta^\mu, z^I$ are local coordinates on $\cE$
one can take the restriction of $x^\mu, z^I$ to $\tilde \cE$ seen as a submanifold in $\cE$ (pullback of $\cE \longrightarrow T[1]\Base$ to the zero section of $T[1]\Base$) as local coordinates on $\tilde \cE$. The above isomorphism can be defined on representatives as follows:
\begin{equation}
    \Upsilon(\dR x^\mu)=\Upsilon(\dR\theta^\mu)=0\,,
    \quad \Upsilon(\theta^\mu)=\dR x^\mu\,,
    \quad \Upsilon(\dR z^I)=\dv z^I\,,
    \quad \Upsilon f(z,x)=f(z,x)\,.
\end{equation}
The properties of $\Upsilon$ are
\begin{equation}
    \Upsilon(\dR\alpha)=\dv \Upsilon(\alpha)\,,
    \qquad
    \Upsilon(L_Q\alpha)=\dh \Upsilon(\alpha)\,,
\end{equation}
so that $\Upsilon\big((\dR+L_Q)\alpha\big) = \dR \Upsilon(\alpha)$.
Moreover, $\Upsilon$ is clearly compatible with the projections
and hence defines the map from $\Functions(T[1]\Base)$
to $\bigwedge^\bullet(\Base)$ which by slight abuse of conventions
we keep denoting by $\Upsilon$.

The degree-preserving sections of $\cE$ are 1:1 with sections of ${\tilde \cE}$ and we define by $\tilde \sigma$ the section of ${\tilde \cE}$ corresponding to $\sigma: T[1]X \to \cE$. Then we have the following diagram:
\begin{equation}
    \begin{tikzcd}[row sep=large]
    \bigwedge^\bullet(\cE)/\cI \arrow[r,"{\Upsilon}"']
    \arrow[d,"\sigma^*_W"']
    & \bigwedge^{(\bullet,\bullet)}({\tilde \cE})
    \arrow[d,"\tilde\sigma^*"']\\
    \Functions(T[1]\Base) \arrow[r,"\Upsilon"']
    & \bigwedge^\bullet(\Base)
    \end{tikzcd}
\end{equation}
It is commutative if $\sigma^*_W$ is precisely the Chern--Weil map 
defined in Section~\ref{sec:multi_AKSZ}. Indeed, it is enough
to check commutativity for functions and basis $1$-forms.
For functions it is obvious while for $1$-forms we get:
\begin{equation}
    \Upsilon\big(\sigma^*_W (\dR z^I)\big)
    =\Upsilon\big(\dx \sigma^*(z^I)
    -\sigma^* (Qz^I)\big) = \dR\tilde\sigma^*(z^I)
    -\dR x^\mu \tilde\sigma^*(D_\mu z^I)\,,
\end{equation}
and
\begin{equation}
    \tilde\sigma^*\big(\Upsilon(\dR z^I)\big)
    =\tilde\sigma^*(\dv z^I)
    =\tilde\sigma^*\big(\dR z^I-\dh z^I)
    =\dR\tilde\sigma^*(z^I)-\dR x^\mu \tilde\sigma^*(D_\mu z^I)\,,
\end{equation}
where we introduced the notation $Q z^I=\theta^\mu D_\mu z^I$.
Note also that $\dh z^I=\dR x^\mu D_\mu z^I$. This diagram shows
that under the identification of differential forms on ${\tilde \cE}$ 
and $\cE$ via $\Upsilon$ the Chern--Weil map is just a pullback
by $\tilde\sigma$.
Note that the Chern--Weil map vanishes on differential forms
from $\cI$ and hence is well-defined on $\bigwedge^\bullet(\cE)/\cI$. 
For instance, $\sigma^*_W(\dR x^\mu)
= \dx \sigma^*(x^\mu)-\sigma^*(Q x^\mu)=0$.

In particular, let $\Omega$ be a $\dR$-closed and $L_Q$-closed
$p$-form on $\cE$ such that $\gh{\Omega}=n-p+1$.
Then $\tilde \Omega \equiv \Upsilon(\Omega)$ is an $(n+1-p,p)$-form 
on ${\tilde \cE}$. Applying $\Upsilon$ to both sides
of $\Omega=(L_Q+\dR)\Theta_W$
one gets $\tilde \Omega=(\dh+\dv) \Upsilon (\Theta_W)$ and the commutativity of the above diagram implies that
\begin{equation}
    \int_\Base \tilde\sigma^* \Upsilon(\Theta_W)
    =\int_{T[1]\Base}\sigma^*_W \Theta_W\,.
\end{equation}
It follows the AKSZ action coincides with its multisymplectic 
analog~\eqref{ms-action} modulo boundary term that vanishes
if $\tilde\Theta$ defined via $\dR\tilde\Theta=\tilde\Omega$
modulo $\dR$-exact form, is taken to be $\Upsilon(\Theta_W)$. Note that if $\Base$ is a boundary of another manifold $\mathscr{Y}$ then the multisymplectic action $\int_\Base \tilde\sigma(\tilde\Theta)$ is a boundary term of $\int_{\mathscr{Y}}\tilde\sigma(\tilde\Omega)$, where it is assumed that $\tilde\sigma$ is extended to a section of the initial bundle extended to $\mathscr{Y}$. This is of course in agreement with the general statement from Section~\ref{sec:multi_AKSZ}.


\section{Conclusions and Discussion}

In this work we explicitly described the multisymplectic generalisation of AKSZ sigma models and presented a variety of examples. Although the main steps of the construction are valid for generic $Q$-bundles, all our examples are background independent and are described by locally trivial $Q$-bundles. The only exception is a version of the construction discussed in Section~\ref{sec:PDE}, where the underlying bundle is a PDE. A full-scale generalisation to the case of generic $Q$-bundles deserves a separate discussion and nontrivial field-theoretical examples. In the case of $p=2$ this generalisation amounts to passing from (presymplectic) AKSZ models to gauge PDEs equipped with compatible presymplectic structures~\cite{Grigoriev:2022zlq}, see also~\cite{Dneprov:2024cvt}.

Although multisymplectic AKSZ models contain the usual AKSZ models as a particular case, they are more naturally viewed as generalisations of presymplectic AKSZ models, because they generally describe theories with local degrees of freedom even if the target space is finite dimensional. This happens because both in the presymplectic and in the multisymplectic setting the Euler–Lagrange equations are consequences of the $Q$-map condition, whereas in the usual AKSZ case these coincide, so that the model is topological unless the target space is infinite dimensional. There is, however, an important difference. A presymplectic BV–AKSZ model, or more generally a presymplectic gauge PDE, naturally encodes not only the action and gauge transformations but also the full local BV formulation of the underlying gauge theory, see~\cite{Grigoriev:2020xec,Grigoriev:2022zlq,Dneprov:2024cvt}. In the multisymplectic case, we generally do not see a natural BV structure encoded in the multisymplectic AKSZ data. Of course, under certain regularity assumptions one can always construct the BV formulation of a given gauge theory, but it is not immediately clear how the BV formulation of a multisymplectic AKSZ model can be directly expressed in terms of its basic structures.

Another open question concerns the multisymplectic analogue of presymplectic gauge PDEs. In the presymplectic setup it is natural and very useful to replace the restrictive condition $Q^2=0$ with a more flexible presymplectic master equation
$\tfrac{1}{2}\imath_Q \imath_Q\omega+Q\mathcal{L}=0$ 
which nevertheless defines a nilpotent differential on a certain subspace and naturally determines the BV formulation~\cite{Grigoriev:2022zlq,Dneprov:2024cvt}.\footnote{Note that in the locally trivial setup the presymplectic master equation takes a more conventional form $\imath_Q \imath_Q\omega=0$, so that it is manifest that it defines a homological vector field on the symplectic quotient of the target space.} Just as in the presymplectic case, such a generalisation is expected to result in a more flexible formalism.

In this work we have limited ourselves to AKSZ-like models whose sources have the form $T[1]\Base$, with $\Base$ being spacetime. Because the AKSZ construction can be  formulated in algebraic terms, one can replace functions on the source (i.e. the exterior algebra of $\Base$) with a more general differential graded algebra that is not necessarily freely generated, while still requiring it to be a module over $\cC^{\infty}(\Base)$ in order not to lose the interpretation in terms of local fields. In this way one can describe nontopological theories even if the target space is a finite-dimensional symplectic $Q$-manifold. For instance, an elegant formulation of Yang–-Mills theory as an AKSZ sigma model of this type was put forward in~\cite{Costello:2011}, see also~\cite{Bonechi:2009kx,Costello:2016mgj,Bonechi:2022aji}. Similar generalisations have recently turned out to be very useful in the AKSZ approach to superfield formulations~\cite{Grigoriev:2025vsl} of supersymmetric systems.

An alternative other generalisation that affects the source space is to replace $T[1]\Base$ with another gauge PDE over $T[1]\Base$ and to relax the defining relations in such a way that the usual conditions are satisfied upon restriction to solutions of the base gauge PDE. This extension precisely corresponds to considering systems coupled to background fields described by the base gauge PDE, see~\cite{Dneprov:2025eoi}. There is little doubt that this generalisation can be carried out in the multisymplectic setting as well, but this is postponed to future work.

As far as the applications are concerned, there are plenty of examples in (mathematical) physics of theories that have higher derivative actions, but this does not always imply that the equations have higher derivatives in time. Examples include higher Chern--Simons theories, Lovelock gravity, Horndeski gravity \cite{Horndeski:1974wa}, Dirac--Born--Infeld Lagrangians, which emerge from string theory \cite{Fradkin:1985qd,Tseytlin:1999dj}, and Galileons, see e.g. \cite{Nicolis:2008in,deRham:2012az}. On the other hand, there are also numerous examples where the kinetic term is naturally of $(d\!\bullet \dots\, d\bullet)$-type and re-expressing the products of the differentials via auxiliary fields does not seem natural. Examples here include the same higher Chern--Simons theories, Lovelock gravity, MacDowell--Mansouri--Stelle--West gravity, Pleba\'nski formulation of gravity \cite{Plebanski:1977zz}, self-dual gravity and its higher-spin extensions, see e.g. \cite{Krasnov:2021nsq,Tran:2022tft, Neiman:2024vit, Mason:2025pbz}. Let us note that at the cost of bringing in additional auxiliary fields one can reformulate multisymplectic theories as presymplectic ones, see e.g. \cite{Dneprov:2024cvt} for an example of Pleba\'nski gravity.

There is also a class of non-gravitational theories, e.g. (self-dual) Yang--Mills theory and its higher-spin extensions \cite{Krasnov:2021nsq,Tran:2022tft,Adamo:2022lah} as well as higher dimensional generalisations of these ideas \cite{Basile:2024raj}, that require a background. The easiest way to embed them into the above setup is to actually couple them to (self-dual) gravity, which makes them of the multisymplectic type if one takes the Pleba\'nski formulation or self-dual gravity, and consider expansion over a particular background.
Note that although (self-dual) Yang--Mills  theory adits a concise presymplectic gauge PDE formulation~\cite{Grigoriev:2022zlq,Dneprov:2024cvt} its coupling to self-dual gravity should be naturally described in the multisymplectic framework, which is also inhomogeneous. Of course, expanding multisymplectic theories over a fixed background would require a generalisation of the multisymplectic formalism to generic $Q$-bundles. 

One immediate application of the formalism introduced in the paper is to higher-spin gravity. As it was discussed in \cite{Sharapov:2021drr}, (chiral) higher-spin gravity admits a presymplectic action that, under certain assumptions (similar to the requirement of non-degenerate vielbein in gravity), is strong enough to recover the complete equations of motion (modulo de Rham cohomology). However, the higher-spin extension of self-dual gravity, which is its subsector, teaches us that the most natural action is of the multisymplectic type. It would be interesting to extend this result to the full theory. 

In addition the equations of motion of chiral higher-spin gravity have \cite{Sharapov:2022nps} the form of those of the Poisson sigma-model \cite{Ikeda:1993fh, Schaller:1994es}.\footnote{Note that Poisson sigma-models are two-dimensional topological theories, but the equations of motion are valid and gauge-invariant in any $d\geq2$. With infinite-dimensional target space PSM' equations can describe propagating degrees of freedom, which is the case in \cite{Sharapov:2022nps}.} Therefore, one can envisage existence of a class of higher dimensional PSM with multisymplectic actions (see also \cite{Chatzistavrakidis:2021nom,Chatzistavrakidis:2022hlu, Chatzistavrakidis:2024utp} for another direction to extend PSM to higher dimensions).

\section*{Acknowledgement}
We are grateful to Euihun Joung and Minkyeu Cho for attracting our attention to the multisymplectic formalism for strings and branes, which was the subject of their contribution in the workshop 
``Conformal higher spins, twistors and boundary calculus'', Mons, July 2025. 
M.G. is indebted to Alexey Kotov for a numerous discussions on $Q$-bundles and Chern--Weil map in particular as well as for collaboration on the related issues. M.G. also acknowledges fruitful discussions with Ivan Dneprov. E.S. is grateful to Andrew Strominger and Lionel Mason for bringing up the Sparling three-form. The work of M.G. was supported  by the ULYSSE Incentive Grant for Mobility in Scientific Research [MISU] F.6003.24, F.R.S.-FNRS, Belgium. The work of T.B. and E.S. was supported by the European Research Council (ERC) under the European Union’s Horizon 2020 research and innovation programme (grant agreement No 101002551). E.S. is a research associate of the Fonds de la Recherche Scientifique – FNRS.

\providecommand{\href}[2]{#2}\begingroup\raggedright\endgroup

\end{document}